\documentclass[fleqn,usenatbib]{mnras}
\usepackage[T1]{fontenc}

\DeclareRobustCommand{\VAN}[3]{#2}
\let\VANthebibliography\thebibliography
\def\thebibliography{\DeclareRobustCommand{\VAN}[3]{##3}\VANthebibliography}

\usepackage{graphicx}	% Including figure files
\usepackage{amsmath}
\usepackage{float}
\usepackage[nice]{nicefrac}
\usepackage{url}

\usepackage{siunitx}
\usepackage{color}
\usepackage{xcolor}
\usepackage{tabularx}

\usepackage{ulem}

\usepackage{newtxtext,newtxmath}

\title[Echelle chemodynamics of CGCG 007-025]
{Shocks and complex chemodynamics in the metal-poor starburst galaxy CGCG\,007-025 revealed through high-resolution echelle spectroscopy}

\author[M. G. del Valle-Espinosa et al.]
{Macarena G. del Valle-Espinosa,$^{1}$\thanks{E-mail: macarena.garciavalle@ed.ac.uk}
Vital Fern{\'a}ndez,$^{2}$\thanks{E-mail: vital.fernandez@userena.cl}
Rub\'en S\'anchez-Janssen,$^{3,1}$
Ricardo Amor{\'i}n,$^{4,5}$ \newauthor
Karla Z. Arellano-C{\'o}rdova,$^{1}$ 
Konstantina Boutsia$^{6,7}$
\\
$^{1}$Institute for Astronomy, University of Edinburgh, Royal Observatory, Edinburgh EH9 3HJ, UK\\
$^{2}$ Michigan Institute for Data Science, University of Michgigan, 500 Church Street, Ann Arbor, MI 48109, US \\
$^{3}$UK Astronomy Technology Centre, Royal Observatory, Blackford Hill, Edinburgh EH9 3HJ, UK\\
$^{4}$Instituto de Astrof\'{i}sica de Andaluc\'{i}a (CSIC), Apartado 3004, 18080 Granada, Spain\\
$^{5}$Centro de Estudios de F\'{\i}sica del Cosmos de Arag\'{o}n (CEFCA), Unidad Asociada al CSIC, Plaza San Juan 1, E--44001 Teruel, Spain\\
$^{6}$Cerro Tololo Inter-American Observatory/NSF NOIRLab, Casilla 603, La Serena, Chile\\
$^{7}$Las Campanas Observatory, Carnegie Observatories, Colina El Pino, Casilla 601, La Serena, Chile\\
}

\date{Accepted XXX. Received YYY; in original form ZZZ}

\pubyear{2023}

\begin{document}
\label{firstpage}
\pagerange{\pageref{firstpage}--\pageref{lastpage}}
\maketitle

% Abstract of the paper
\begin{abstract}
% This is a simple template for authors to write new MNRAS papers.
% The abstract should briefly describe the aims, methods, and main results of the paper.
% It should be a single paragraph not more than 250 words (200 words for Letters).
% No references should appear in the abstract.

%Nearby metal-poor starburst dwarf galaxies present a unique opportunity to probe the physics of high-density star formation with a detail and sensitivity unmatched by any observation of the high-z Universe. These chemically unevolved galaxies also offer insights into the synthesis, dispersal, and ejection of metals in galaxies, from the inflows of minimally processed material to the metal-enriched outflows driven by intense star formation events.
\noindent We use Magellan/MIKE echelle spectroscopy to conduct an in-depth chemodynamical analysis of the most luminous star-forming region within the metal-poor starburst dwarf galaxy CGCG 007-025. 
Leveraging the exceptional high resolution (R$\sim$50,000) and broad wavelength coverage, we apply Bayesian inference to simultaneously model the fluxes of 30 emission lines spanning the wavelength range 3400-9200\AA. 
Employing a two-region ionisation model, we characterise various gas properties including electron temperature, electron density, and chemical abundances across different elements. 
Our direct-method inferred metallicity yields $\rm 12+\log(O/H)=7.77\pm0.03$, placing the galaxy in the metal-poor regime. Furthermore, Metal-to-Oxygen ratios such as log(S/O), log(Ne/O) or log(Ar/O) are in full agreement with the values derived for the Milky Way, consistent with expectations from stellar evolutionary models. 
The brightest emission lines are kinematically complex, with modelling requiring up to four distinct components.
The exceptional resolution and signal-to-noise ratio of the data unveil asymmetric and wide ($\sigma_{\ion{He}{ii}} \approx$ 35km/s) \ion{He}{ii}$\lambda$4686 emission. The flux ratio of this nebular line, together with the absence of other high ionisation species such as [\ion{Ne}{v}]$\lambda$3426, indicates the presence of fast radiative shocks. %This echelle dataset allow us to explore SF regions i is the first chemical study using echelle spectroscopy made on a star forming region outside of the Local Group. evidences the power of echelle spectroscopy to provide detail chemodinamical analysis of SF regions in the Local Volume.
This dataset underscores the capability of echelle spectroscopy in delivering comprehensive chemodynamical analyses of starbursts in the Local Volume. 
\end{abstract}

% Select between one and six entries from the list of approved keywords.
% Don't make up new ones.
%%%%%%%%%%%%%%%%%%%%%%%%%%%%%%%%%%%%%%%%%%%%%%%%%%

%%%%%%%%%%%%%%%%% ALIASES %%%%%%%%%%%%%%%%%%%%%%%%

\defcitealias{jr:fernandez2023}{Paper I}
\defcitealias{jr:mgve2023}{Paper II}
\definecolor{corrections}{HTML}{000000}
\definecolor{corrections3}{HTML}{000000}

%%%%%%%%%%%%%%%%%%%%%%%%%%%%%%%%%%%%%%%%%%%%%%%%%%

%%%%%%%%%%%%%%%%% BODY OF PAPER %%%%%%%%%%%%%%%%%%
\section{Introduction}
\label{sec:Introduction}

Nearby interacting low-mass galaxies stand out as the most ideal laboratories to study the conditions of star-formation at high-redshift \citep[e.g., ][]{jr:PapaderosOstlin2012}, as they do not only share comparable stellar masses, metallicities and specific star-formation rates (SFR) with the average high-$z$ galaxy \citep{jr:wiesz2011,jr:izotov2021}, but also their interstellar medium (ISM) properties and environmental conditions resemble the ones of the high-$z$ galaxy population \citep[e.g., ][]{jr:bradford2015}.
First, their low metallicities together with their high gas mass fractions \citep{jr:HaywardHopkins2017} mimic the turbulent ISM and therefore bursty star-formation histories (SFHs) and feedback intensities of their high-$z$ counterparts \citep{jr:muratov2015,jr:trebitsch2017}.
Second, their tidal interactions not only foster starbursting episodes but also canalise large quantities of preexisting metal-poor gas from the outskirts towards the central regions \citep[e.g., ][]{jr:luo2014}.

With the advent of the \textit{James Webb Space Telescope} (JWST), the characterisation of %the MZR
low mass, star-forming galaxies has been possible for the first time at $z>$ 3 \citep[e.g., ][]{jr:curti2023arXiv230408516C,jr:nakajima2023arXiv230112825N,jr:sanders2023arXiv230308149S}. 
The very first NIRSpec/JWST observations, using the Early Release Observations (EROs) of the galaxy cluster SMACS J0723.3-7327, revealed three emission lines galaxies at $z\sim$ 8 whose emission lines ratios resemble local EELGs \citep[extreme emission line galaxies, e.g., ][]{jr:schaerer2022A&A...665L...4S}. Expectedly, their metallicities (7.0 $<$12+log(O/H) $<$ 8.1) and stellar masses (7$<$logM$\star<$9) also endorse the properties of prototypical local metal-poor starburst dwarf galaxy \citep[e.g.][among others]{jr:arellano-cordova2022ApJ...940L..23A,jr:curti2023MNRAS.518..425C,jr:trump2023ApJ...945...35T}. 
In the framework of optical line diagnostics, several studies followed the former pioneering works \textcolor{corrections}{with JWST in the characterisation of high redshift ($z>$ 6)} emission line galaxies. For example, \citet{jr:cameron2023A&A...677A.115C} found that the locus of the R2-R3 diagram (i.e., [\ion{O}{iii}]$\lambda$5007/H$\beta$ vs. [\ion{O}{ii}]$\lambda\lambda$3726,3729/H$\beta$ ) for galaxies at $z>$ 5 was in good agreement with the extreme $z\sim$ 0 dwarf starbursts classified as Green Peas \citep{jr:Cardamone2009MNRAS.399.1191C,jr:yang2017ApJ...844..171Y,jr:amorin2010ApJ...715L.128A,jr:amorin2012ApJ...749..185A,jr:fernandez2022MNRAS.511.2515F} and Blueberries \citep{jr:yang2017ApJ...847...38Y}. Moreover, \citet{jr:curti2023arXiv230408516C} studied the low-mass end of the Mass-Metallicity Relation \citep[MZR, ][]{jr:tremonti2004,jr:sanders2021ApJ...914...19S} of galaxies at 3 $<z<$ 10 and found a remarkable agreement with the slope of the MZR at these redshifts and the one from local EELGs. %\textcolor{red}{try to search for more examples?}
However, the normalisation of the MZR of high redshift galaxies is more debatable. For example, \citet{jr:Brinchmann2023MNRAS.525.2087B} reported accurate masses and metallicities which laid above the ones observed in low-$z$ Green Peas. %The author concluded that these high redshift galaxies are still above the mass and metallicities observed in Green Peas. %As predicted by \textcolor{red}{Amorin 2017} fig 4 , these galaxies experience a large mass gain while keeping a flat metallicity gradient. 

\textcolor{corrections}{All these measurements lead to the following conclusion: There are still galaxies in the local Universe with nebular properties closely resembling those in the 1 Gyr old universe. Yet, JWST studies are commonly restricted to the use of bright emission lines due to signal-to-noise limitations. Moreover, high-$z$ integrated spectra do not provide any spatial information, crucial to study the interplay between the ionising sources and their surrounding medium. Once again, galaxies in the local Universe fulfil both the brightness and size requirements to overcome these barriers. }
The study of nearby, chemically young high-ionisation dwarf galaxies can help to constrain and interpret the physical properties of the high redshift counterparts \citep[e.g., ][]{jr:Senchyna2017MNRAS.472.2608S, jr:senchyna2019MNRAS.488.3492S,jr:berg2021ApJ...922..170B}, such as dust content, electron density and temperature structure, gas-phase metallicity, and the ionization state of the gas \citep[e.g., ][among others]{jr:IzotovThuan1999ApJ...511..639I, jr:SanchezAlmeida2016ApJ...819..110S, jr:berg2019ApJ...874...93B, jr:Mingozzi2022ApJ...939..110M}.

At a distance of $d_L = 23 \pm 5$ Mpc \citep{jr:kourkchi2020} and with a stellar mass of ${\rm M_{\star} = 1.2 \times 10^8 ~M_\odot}$ \citep{jr:marasco2022}, the starbursting dwarf galaxy CGCG 007-025 satisfies all the criteria to be treated as a local analogue of high redshift galaxies. 
This galaxy was first named from Catalogue of Galaxies and of Clusters of Galaxies \citep[see][]{jr:ZwickyKowal1968cgcg.bookR....Z} and SHOC270 from the SDSS H II-galaxies with Oxygen abundances Catalog \citep[see ][]{jr:Kniazev2004ApJS..153..429K}, and it is also known as J094401.86-003832.1 \citep[from SDSS,][]{jr:cgcgSDSS}, MCG +00-25-010 \citep[from the \textit{Millennium Galaxy Catalogue},][]{jr:cgcgMGC} and SB2 in \citet{jr:shirazi2012MNRAS.421.1043S,jr:Senchyna2017MNRAS.472.2608S,jr:Senchyna2022} sample.

Using echelle spectroscopy of the brightest SF region in CGCG 007-025 we provide a complete view of the kinematics and chemical structure using the whole set of optical lines available from 3350\AA~to 9410\AA~with a resolution of R$\sim$40,000. 
This paper is a follow up analysis on the results presented in \citet{jr:fernandez2023} and \citet{jr:mgve2023} (hereafter, \citetalias{jr:fernandez2023} and \citetalias{jr:mgve2023}). 

In Section \ref{sec:Data} we describe the observations and data reduction, the methodology is presented in Section \ref{sec:Methods}, the main results from the kinematics of the lines as well as the derived chemistry are compiled in Section \ref{sec:Results} with the discussion in Section \ref{sec:Discussion}. We summarise our findings in Section \ref{sec:Conclusions}.

\section{Data}
\label{sec:Data}

%===========================================================================================

\subsection{Observational data}

The observations consist in echelle spectroscopy of the brightest star forming region in the dwarf starbursting galaxy CGCG 007-025 identified in \citetalias{jr:mgve2023}.
The observations were carried out using the instrument MIKE \citep[Magellan Inamori Kyocera Echelle, ][]{jr:MIKE2003SPIE.4841.1694}, a echelle-type spectrograph installed at the 6.5m Magellan 2- Clay telescope. Its two arms operate simultaneously, giving spectral coverage from 3350 to 5060 \AA~in the blue arm and from 4830 to 9410 \AA~in the red. Among the different slit sizes available, we selected a slit of $0.7\arcsec\times~ 5\arcsec$ (dispersion $\times$ spatial directions). We measured the spectral resolution in each arm using the Th-Ar lines from the arc observations. The spectral resolution, defined as the FWHM of the arc lines, varied from 0.09 to 0.12 \AA~(about 7.8 ${\rm km s^{-1}}$) in the blue, and from 0.16 to 0.24 \AA~(about 9.1 ${\rm km s^{-1}}$) in the red.

The observations were carried out in three different nights, with a total time on source of 3.6 hours. In Table \ref{tab:nightlog} we presented the observing log, where we list the observing date, the exposure times and the airmass for each night. Although the Atmospheric Dispersal Corrector (ADC) was not operational for the last two nights, the observations were made with an airmass value lower than 1.6, reducing the impact of the atmospheric dispersion to negligible levels.

\begin{table}
    \centering
    \resizebox{0.48\textwidth}{!}{  
    \begin{tabular}{ccccc}
        Region & Date & Airmass & Exposure time & ADC \\ \hline
        Brightest clump & 17 April 2021 & 1.158 & 3x1200 & Yes \\
        Brightest clump & 30 May 2021 & 1.366 & 3x1200 & N/A \\
        Brightest clump & 31 May 2021 & 1.220 & 5x1200 & N/A \\
    \end{tabular}
    }
    \caption{Night log of the different observing nights}
    \label{tab:nightlog}
\end{table}

\begin{figure}
    \centering
    \includegraphics[trim=20 20 80 60,clip, width=0.45\textwidth]{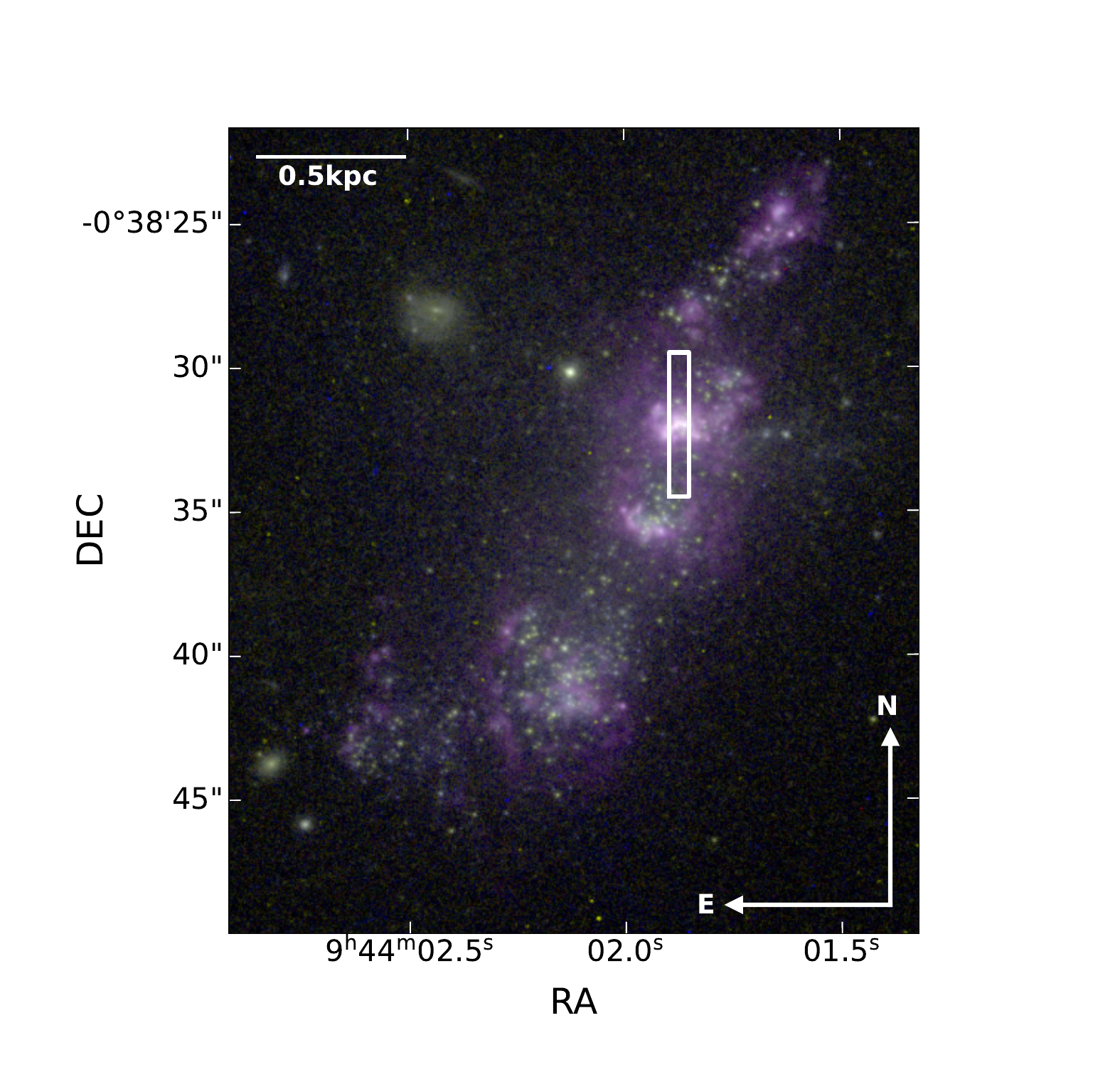}
    \caption{HST/WFC3 colour image created with the F275W, F336W, F435W, F606W, F657N, F875W filters. The white box indicates the location of the slit.}
    \label{fig:MIKEslit}
\end{figure}

%===========================================================================================

\subsection{Data reduction}
\label{sec:DataReduction}

The echelle data of the three nights were reduced using the \textsc{mike pipeline}\footnote{\url{https://code.obs.carnegiescience.edu/mike}} available at the Carnegie Python Distribution \citep[CarPy, ][]{jr:CarPy2000ApJ...531..159K,jr:CarPy2003PASP..115..688K}. Briefly, the pipeline follows standard reduction procedures for each arm independently to produce 1D wavelength calibrated spectra for every detected order. Bias images taken during the observing nights are combined and subtracted from all remaining frames. Then, the pixel-to-pixel variations across the CCD are corrected using the averaged and normalised flat-field frames. The limits of each order are calculated using the Th-Ar arcs, and the order curvature is corrected using a quartz lamp spectrum. Once the 2D distortions are corrected, the pipeline extracts 1D spectra for each order in both the arc and the science frames. The 1D arc spectra is lastly used for the wavelength calibration of the science frames. The same procedure was applied for the reduction of the selected spectrophotometric standard stars \citep[Hiltner 600 for the first night and LTT 3864 for the others, ][]{jr:standardStars1994PASP..106..566H}. Using the extracted and wavelength-calibrated spectra of the standard stars we calibrate in flux our observations of the SF region.

Figure \ref{fig:MIKEslit} shows the location of the slit during the three observing nights. Since the slit does not cover any sky area and we do not have an sky-only pointing, the final spectra is not corrected from telluric contamination. However, the high spectral resolution allow us to deblend sky emission lines from nebular ones.
A compilation of all the detected emission lines with amplitude-over-noise (AON) higher than 5 is displayed in Figure \ref{fig:emissionLines}.

\section{Methodology}
\label{sec:Methods}

\subsection{Emission line modelling}

Our emission line modelling is performed locally in each of the orders to avoid problems in the continuum determination. For most of the cases, the model consists of a combination of a Gaussian profile describing the emission line of interest with a one degree polynomial to account for the continuum contribution. If one (or several) sky lines fall in the fitting window, we include them in the model routine as Gaussian profiles with a width equal to the resolution in that order \textcolor{corrections}{(typically $\sim$8 km/s). We use the UVES atlas of sky emission lines\footnote{\url{https://www.eso.org/observing/dfo/quality/UVES/pipeline/sky_spectrum.html}} for the sky contamination modelling, which catalogues sky emission lines from 3140 to 10430 Å at a similar resolving power as the echelle data presented in this work. Note the width of the science lines is double this value, meaning we can model and remove sky contamination confidently}. Since we do not include a telluric absorption correction, lines falling in the wavelength ranges 6850-7000\AA~ and 7580-7700 \AA~\citep[the regions strongly affected by the atmospheric O$_2$ band, ][]{jr:stevenson1994MNRAS.267..904S} are excluded from our analysis.

The brightest line profiles in this echelle spectrum required three kinematically different Gaussian components: a narrow profile with velocity dispersion $\sigma_{\text{narrow}} \sim 15$km/s, an intermediate profile with $\sigma_{\text{medium}} \sim 40$km/s and a broad line with $\sigma_{\text{broad}} \sim 200$km/s. 

In \citetalias{jr:mgve2023}, we modelled the line profiles of CGCG 007-025 using MUSE IFU data with an intermediate resolving power of R$\sim$2000. The MUSE data also revealed a complex line profile in H$\alpha$, where three components were needed to describe the core and wings of the line: an unresolved ($\sigma<50$km/s) narrow component, an intermediate component with velocity dispersion of $\sigma\sim 150$km/s and a broad component with $\sigma\sim1000$km/s. 
This suggests that the broad component detected in MIKE is actually the intermediate component detected in MUSE, with the unresolved narrow Gaussian in MUSE being resolved into two different components. The MUSE broad component is so dispersed at the echelle resolution that it could not be detected. A more detailed analysis on the different components model is presented in Section \ref{sec:multipleLines}.

\subsection{Gas phase physical properties and direct method metallicities}

In the ideal scenario, the chemical composition of the star-forming gas is measured directly from the quantified emission of all the ionised species for the considered elements. This observation-based approach is known as the direct method. This methodology requires high S/N spectra to observe the electron density $(n_e)$ and temperature $(T_e)$ sensitive, auroral lines. In this model, the observed line fluxes are parameterised relative to $H\beta$ as:

\begin{equation}
\frac{F_{X^{i+},\,\lambda}}{F_{H\beta}}=X^{i+}\frac{\epsilon_{X^{i+},\,\lambda}\left(T_{e},\,n_{e}\right)}{\epsilon_{H\beta}\left(T_{e},\,n_{e}\right)}\cdot10^{-c\left(H\beta\right)\cdot f_{\lambda}}\label{eq:fluxFormula}
\end{equation}

where $\nicefrac{\epsilon_{X^{i+},\,\lambda}}{\epsilon_{H\beta}}$ is the relative transition emissivity at wavelength $\lambda$, for an ion with abundance $X^{i+}$. $c\left(H\beta\right)$ is the relative logarithmic extinction coefficient and $f_{\lambda}$ is the reddening curve. Since the ionised species commonly observed in the interstellar medium have distinctive ionisation potential energies, the model in eq. \ref{eq:fluxFormula} assumes that ions have a characteristic temperature and density. In practice, this is simplified to several high-to-low "ionisation zones", each with a different electron temperature. In contrast, at the low densities encountered in star-forming regions, this parameter has a low impact on the photons emission from these transitions. Consequently, it is acceptable to use a uniform density across the ionisation regions.

As we did in \citetalias{jr:fernandez2023}, we use a Bayesian sampler based on neural networks to explore the chemical parameter space. This model was presented in \cite{jr:fernandez2019MNRAS.487.3221F} against the traditional direct method workflow. Since this sampler can fit the set of eqs.\ref{eq:fluxFormula} for all the transitions observed, it provides a more reliable uncertainty of the measurements. Unlike in \citetalias{jr:fernandez2023}, the MIKE wavelength range, provides access to the $[OIII]4363$\AA{} auroral line. This enables us to model two regions: a low ionisation zone characterised by $T_{low} = T[SIII]$ for the $O^+$, $N^+$, $S^+$, $S^{2+}$, $Ar^{2+}$ ions and a high ionisation zone with $T_{high} = T[OIII]$ for $y^+$, $O^{2+}$ and $Ar^{3+}$. Both zones assume the same electron density, but since the sampler fits all lines simultaneously this parameter measurement combines the prediction from the [\ion{S}{ii}]$\lambda\lambda$6716,6731\AA{} and [\ion{O}{ii}]$\lambda\lambda$3726,3729\AA{} doublets.

In \citetalias{jr:fernandez2023}, we evaluated how the uneven relative uncertainty between the auroral and nebular lines affects the accuracy and precision of the measurements. However, thanks to the high S/N of the present observations, both line types have relative uncertainties within the same order. The only precaution necessary was to make sure that all the lines considered belonged to the same kinematic component.

\section{Results}
\label{sec:Results}

\subsection{Kinematics of the ionised gas}
\label{sec:multipleLines}

\begin{figure}
    \centering
    \includegraphics[width=0.48\textwidth]{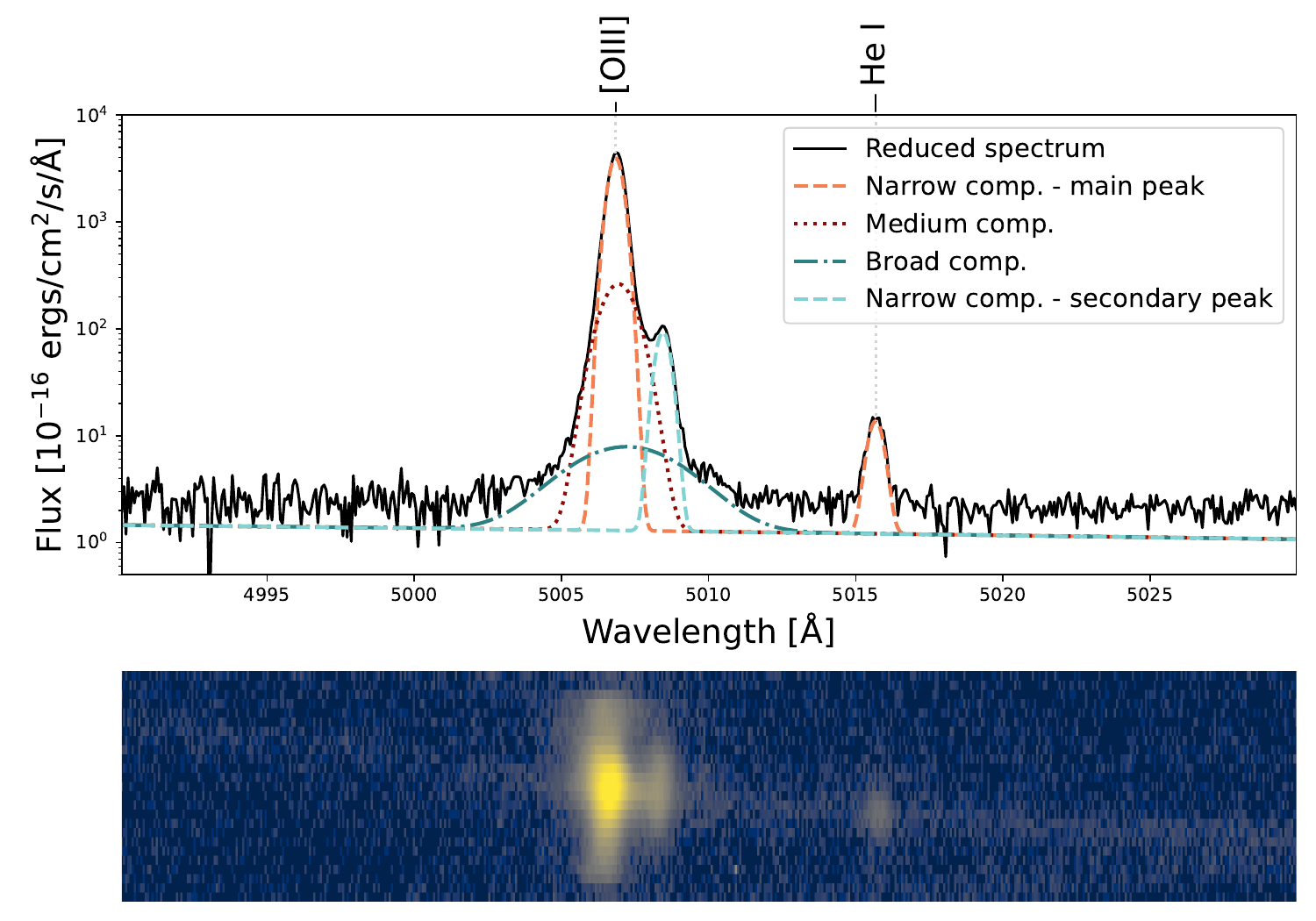}
    \caption{\textit{Top panel}: MIKE spectrum (black-solid line) at the location of the emission lines [\ion{O}{iii}]$\lambda$5007 and \ion{He}{i}$\lambda$5015. The location of both lines is marked in the plot. The narrow component for these emission lines is displayed with an orange-dashed line. In the case of [\ion{O}{iii}]$\lambda$5007, a secondary narrow peak redshifted from the main peak is visible. \textit{Bottom panel}: raw 2D echellogram zoomed in around the [\ion{O}{iii}]$\lambda$5007 order. As seen in this plot, the emission from the secondary peak is coming from the same line of sight as the main peak. Notice the trace of the order is tilled since we are displaying the raw data.}
    \label{fig:O3_doublePeak}
\end{figure}

The echelle spectrum gives us access to emission lines in the wavelength range from 3300 to 9100 \AA. We are able to measure a total of 77 emission lines with a signal to noise higher than 5. In most of the cases the line profiles are well described by a single narrow component, but for lines with AON $>$ 50 an intermediate component is required to model the wings of the line. Moreover, for the four brightest emission lines in the spectrum (i.e., H$\alpha$, [\ion{O}{iii}]$\lambda\lambda$4959,5007 and H$\beta$) we detect the presence of a secondary narrow component redshifted from the main peak. Figure \ref{fig:O3_doublePeak} displays the complexity of the [\ion{O}{iii}]$\lambda$5007 line profile, where we need up to four Gaussian components to described the overall shape of the line: three components to describe the core and wings of the highest intensity, plus a fourth narrow component to model the secondary, redshifted line.  As can be seen in the raw 2D echellogram (Fig. \ref{fig:O3_doublePeak}) this latter emission originates purely along the line of sight of the star-forming-\textcolor{corrections}{knot}, i.e., it's spatially indistinguishable from the brighter emission line.
The best fit parameters for all the lines present in our data are collected in Tables \ref{tab:lineFitsBLUE} and \ref{tab:lineFitsRED}.

\begin{figure}
    \centering
    \includegraphics[width=0.48\textwidth]{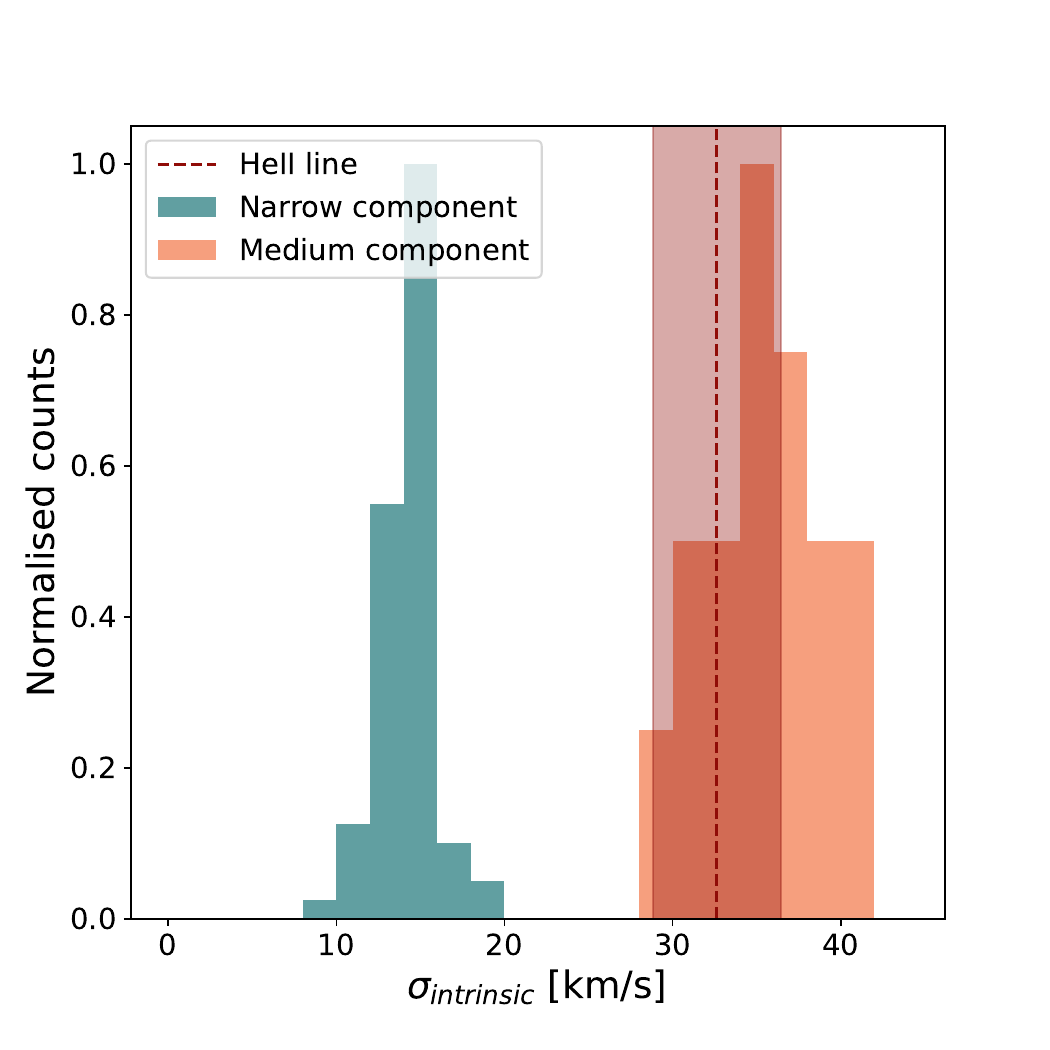}
    \caption{Density histrograms showing the velocity dispersion values, corrected from instrumental and thermal broadening, of the modelled emission lines --both, recombination and forbidden. The teal histogram correspond to the velocity dispersion of the narrow component, while the intermediate component distribution is displayed in orange. The brown dashed line marks the velocity dispersion of the \ion{He}{ii}$\lambda$4686 emission line, well modelled with a single Gaussian with $\sigma=33\pm3$km/s and fully compatible with the velocity dispersion of the intermediate Gaussian component of the other elements.}
    \label{fig:Sigmas}
\end{figure}

In Figure \ref{fig:Sigmas} we show the normalised distribution of the velocity dispersion values for the narrow and medium components of the modelled lines. These velocity dispersion values have been corrected from instrumental broadening --using the appropriate value associated to each order derived from the calibration arcs-- as well as from thermal broadening --using the doppler correction $\sigma_T=\sqrt{k \cdot T/m}$--. All the narrow components (teal histogram) conglomerate around $\sigma_\text{narrow} = 14 \pm 2$ km/s, while the medium component (orange histogram) present a median velocity dispersion of $\sigma_\text{medium} = 37 \pm 9$ km/s. 

\begin{figure*}
    \centering
    \includegraphics[width=0.8\textwidth]{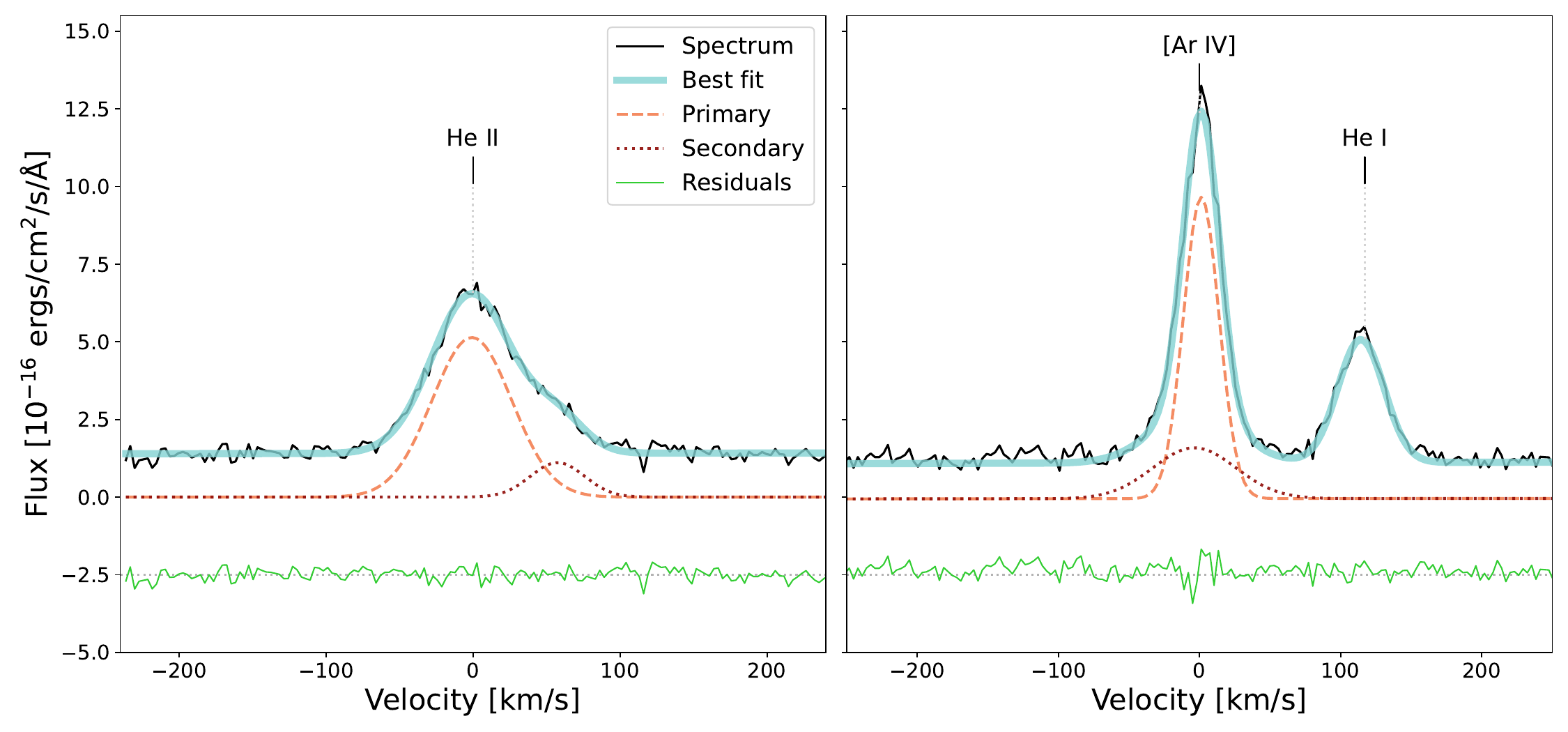}
    \caption{\textcolor{corrections}{Zoom around the \ion{He}{ii}$\lambda$4686, [\ion{Ar}{iv}]$\lambda$4711 and \ion{He}{i}$\lambda$4713 lines. Note that the \ion{He}{ii} primary component, modelling the core of the line, is broader than the other nearby emission lines in the same echelle order (such as [\ion{Ar}{iv}]). Moreover, the HeII displays an asymmetry towards the red. This excess of emission can be modelled with a gaussian component, however its kinematical properties can not be associated with any other line.}} %(grey-shaded area) was masked during the fitting process to better constrain the centre of the line, but its broadness is not affected by the masking. %Anadir comentario de asimetria en HeII, masking al ala roja, y fit cambia si no enmascaras pero sigue siendo ancho
    %}
    \label{fig:HeIIbroad}
\end{figure*}

Remarkably, the \textcolor{corrections}{core of the} \ion{He}{ii}$\lambda$4686 is best modelled \textcolor{corrections}{with a} Gaussian with velocity dispersion of $\sigma=33\pm3$ km/s. In Figure \ref{fig:Sigmas}, we marked this value using a brown dashed line, whose location is consistent with the velocity dispersion distribution associated to the medium component of the rest of the elements. 
CGCG 007-025 is part of the COS Legacy Archive Spectroscopy Survey \citep[CLASSY, ][]{jr:berg2022}, a treasury survey devoted to the study of systems with properties resembling to galaxies at the re-ionisation era.
Figure 3 from \citet{jr:arellano-cordova2022ApJ...935...74A} compiles the archival spectra of CGCG 007-025 (there named J0944-0038) from 3 different instruments with R$<$4500. At these resolutions, the \ion{He}{ii} does not display a different velocity dispersion than the rest of the lines. As a sanity check, in Figure \ref{fig:HeIIbroad} we compare the \ion{He}{ii} line with the adjacent lines in the same spectral order: [\ion{Ar}{iv}]$\lambda$4711 and \ion{He}{i}$\lambda$4713. In this figure the broader line profile of \ion{He}{ii} is noticeable, with an asymmetry towards the red part. \textcolor{corrections}{This excess of emission towards the red wing of the line can be modelled with a Gaussian redshifted $\sim$55 km/s of the core and with $\sigma\approx 20 \pm 2$ km/s. These kinematical features do not agree with any other line seeing in this spectrum, leaving the origin of the asymmetry on the \ion{He}{II} as unknown. Going back to the core of the \ion{He}{II} emission, its comparison with the core of [\ion{Ar}{iv}]} rules out the broadening of the \ion{He}{ii} line as an instrumental effect, \textit{placing} its origin in physical processes within the ionised region and only observable when high spectral resolutions (R$\sim$ 40,000) are used. 

\subsection{Physical conditions of the SF region}

\renewcommand{\arraystretch}{1.5}
\begin{table}%[h]
    \centering
    \begin{tabular}{ccc} \hline
        Parameter                       & This work (MIKE/Magellan) & CLASSY (MODS/LBT)      \\ \hline
        $n_e (cm^{-3})$                 & $250^{60}_{30}$           & $120 \pm 30$      \\
        $T_{low} (K)$                   & $15100^{700}_{500}$       & $15.100 \pm 900$  \\
        $T_{high} (K)$                  & $15900^{500}_{200}$       & $15.500 \pm 300$  \\
        c($H\beta$)                     & $0.25^{0.08}_{0.07}$      &                   \\
    \end{tabular}
    \caption{Physical properties for the brightest star forming region in CGCG 007-025. For comparison, the reported values from \citet{jr:arellano-cordova2022ApJ...935...74A} for the CLASSY collaboration are included in the table. The tabulated values represent the median of the distribution, with the 16$^{\text{th}}$ (84$^{\text{th}}$) percentile being the lower (upper) error.}
    \label{tab:PhysicalProperties}
\end{table}
\renewcommand{\arraystretch}{1}

Table \ref{tab:PhysicalProperties} summarises our results on the physical properties of the star-forming region derived \textcolor{corrections3}{from the narrow component}, as well as the same quantities as measured by the CLASSY collaboration. 
Our extinction coefficient was derived using the ratio between the Paschen and Balmer with respect to H$\beta$. We measured a value of c(H$\beta$) = $0.25^{0.08}_{0.07}$, fully consistent with the coefficient we obtained using the MUSE data in \citetalias{jr:fernandez2023} of c(H$\beta$) = $0.27^{0.07}_{0.08}$.

The electron density of the brightest star-forming region in CGCG 007-025 is $n_e = 250^{60}_{30}$. In \citetalias{jr:fernandez2023} we found this region presents a density gradient, with peak value of $n_e = 378^{34}_{63}$ at the centre of the region and decreasing towards the outskirts until $n_e=177^{84}_{64}$. Since the slit of the echelle observations is covering the entirety of the star-forming clump, it is not surprising the electron density we recover is in between these two values.
In addition to the [\ion{O}{ii}] and [\ion{S}{ii] } density diagnostics, we have also calculated the electron density using the [\ion{Cl}{iii}]~$\lambda$5518/$\lambda$5538 and [\ion{Ar}{iii}]~$\lambda$4711/$\lambda$4741 ratios. These diagnostics represents the intermediate and high ionization density structure of the gas. For CGCG007-025, we report $n_{\rm e}$[\ion{Cl}{iii}] = 500$\pm$700 cm$^{-3}$  and  $n_{\rm e}$[\ion{Ar}{iv}] = 1800$\pm$400 cm$^{-3}$. Our results imply a gradient in the density of CGCG007-025 ranging between 250 cm$^{-3}$ to 1800 cm$^{-3}$ (see Table~\ref{tab:PhysicalProperties}). These results are in agreement with those reported in \citet{jr:Mingozzi2022ApJ...939..110M} within the uncertainties. Since the [\ion{Cl}{iii}] lines are usually faint and the [\ion{Ar}{iv}] lines are mostly blended with He I in low-resolution spectra, we have discarded such diagnostics for the analysis of chemical abundances to better compare with results reported in the literature. However, the inclusion of the density diagnostic of [\ion{Ar}{iv}] does not change the results concerning the physical conditions and chemical abundances.

For the low ionisation species, we measure a temperature of $T_{low} = 15100^{700}_{500}$ K. For this same region, in \citetalias{jr:fernandez2023} we reported a $T_{low} = 15030^{1192}_{643}$ K, consistent with the results presented in this paper. We recover a $T_e$ for the high ionisation of $T_{high} = 15900^{500}_{200}$ K. 
As part of the CLASSY analysis, 
\citet{jr:arellano-cordova2022ApJ...935...74A} derived $T_{\rm e}$ for the brightest SF of CGCG 007-025 using a set of observations with different aperture sizes. In order to reduce the impact of aperture effects, we will only consider the results using the MODS/LBT spectrum of CGCG 007-025. The data from MODS/LBT matches the slit position and width as our echelle configuration. \citet{jr:arellano-cordova2022ApJ...935...74A} measured $T_{[OII]} = 15100\pm900$ K for the low ionisation temperature and for the high ionisation $T_{[OIII]} = 15500\pm300$ K. Both of these values are in excellent agreement with our measurements.

\renewcommand{\arraystretch}{1.5}
\begin{table}%[h]
    \centering
    \begin{tabular}{cc} \hline
        Ionic abundance                 & This work (MIKE/Magellan)     \\ \hline
        $y^+$                           & $0.070^{0.002}_{0.003}$  \\
        {\tiny $\dfrac{O^+}{H^+}$}      & $6.74^{0.14}_{0.10}$     \\
        {\tiny $\dfrac{O^{2+}}{H^+}$}   & $7.73^{0.02}_{0.03}$     \\
        {\tiny $\dfrac{N^+}{H^+}$}      & $5.38^{0.03}_{0.03}$     \\
        {\tiny $\dfrac{S^+}{H^+}$}      & $5.09^{0.03}_{0.05}$     \\
        {\tiny $\dfrac{S^{2+}}{H^+}$}   & $5.88^{0.04}_{0.04}$     \\
        {\tiny $\dfrac{Ar^{2+}}{H^+}$}  & $5.27^{0.03}_{0.03}$     \\
        {\tiny $\dfrac{Ar^{3+}}{H^+}$}  & $4.91^{0.02}_{0.02}$     \\
        {\tiny $\dfrac{Ne^{2+}}{H^+}$}  & $7.03^{0.02}_{0.02}$     \\
        {\tiny $\dfrac{Fe^{2+}}{H^+}$}  & $4.64^{0.03}_{0.03}$     \\
    \end{tabular}
    \caption{Chemical abundances for the brightest star forming region in CGCG 007-025. The tabulated values represent the median of the distribution, with the 16$^{\text{th}}$ (84$^{\text{th}}$) percentile being the lower (upper) error. The metal abundances are in the form $12 + \log(X/H)$.}
    \label{tab:ChemicalAbundances}
\end{table}
\renewcommand{\arraystretch}{1}

Table \ref{tab:ChemicalAbundances} compiles the ionic abundances of our model. Using the ionic abundances for $O^+$ and $O^{2+}$, and considering they are the main components \citep{jr:Pagel1978MNRAS.184..569P}, we determine the gas phase metallicity to be $12+log(O/H)= 7.77\pm0.03$. These results are compatible within the errors with the values reported in \citet{jr:arellano-cordova2022ApJ...935...74A}. For the sulphur ions, there is a non negligible component coming from the $S^{3+}$ ion that needs to be taken in to account when estimating the total sulphur abundance. In \citet{jr:Fernandez2018MNRAS.478.5301F} a calibration for the ionisation correction factor (ICF) of the $S^{3+}$ was given by 
\begin{equation}
    \text{log} \left(\dfrac{Ar^{2+}}{Ar^{3+}}\right) = a \cdot \text{log} \left(\dfrac{S^{2+}}{S^{3+}}\right) + b
\end{equation}

with $a = 1.162\pm0.006$ and $b=0.05\pm0.01$. We obtain an ICF($S^{3+}$) = 1.47$\pm$0.20, which results in a total sulphur abundance of $log(S/H) = 6.11^{0.09}_{0.10}$. Our ionic and total abundance for the sulphur from the echelle data are also in good agreement with the values reported in \citetalias{jr:fernandez2023}.

We can also calculate the total abundance of argon and neon. To account for the unobserved ions of these species, we use the ICFs derived by \citet{jr:amayo2021MNRAS.505.2361A}. These ICFs have been calibrated using photoionisation models and constraints in log(O/H), log(N/O) and the log([\ion{O}{iii}]/H$\beta$) vs. log([\ion{N}{ii}]/H$\alpha$) BPT diagram. All of these constraints are satisfied by the SF region under study. For argon, the corresponding factor is ICF($Ar^{2+} + Ar^{3+}$) = 0.9$\pm$0.1, yielding to a total abundance of log(Ar/H) = 5.42$\pm$0.04. For neon, we obtain ICF($Ne^{2+}$) = 1.04$\pm$0.1 and a total abundance of log(Ne/H) = 7.07$\pm$0.05.

\section{Discussion}
\label{sec:Discussion}

\subsection{The absence  of extremely high ionised lines}

The presence of ionised species with ionisation potential (IP) above 70 eV  in HII regions cannot be explained by pure stellar photo-ionisation \citep[e.g., ][]{jr:berg2021ApJ...922..170B,jr:Olivier2022ApJ...938...16O}, because the ionising spectra of the most massive stars is not hard enough to produce the large amounts of ionising photons capable of generating these high ionisation transitions \citep[][]{jr:stasinska2015A&A...576A..83S,jr:gutkin2016MNRAS.462.1757G}. Therefore, other exotic scenarios have been invoked to produce such lines, such as the presence of Ultra Luminous X-ray sources and High Mass X-ray Binaries \citep[ULX and HMXB, ][]{jr:schaerer2019A&A...622L..10S,jr:simmonds2021A&A...656A.127S}, shocks \citep[e.g.,][]{AM19} or active galactic nuclei \citep[AGNs, ][]{jr:feltre2016MNRAS.456.3354F}. 

According to the unified model of AGNs, the relativistic jets in contact with the torus surrounding the Super Massive Black Hole (SMBH) both provide the energy and density necessary to highly ionised metallic species \citep{2006agna.book.....O}. As such, high IP lines like [\ion{Ne}{v}]$\lambda$3426 in the optical \citep{jr:mingoli2013A&A...556A..29M} and other ratios of infrared coronal lines such as  [\ion{Fe}{xiii}]/[\ion{Fe}{vi}] at $1\mu$m or [\ion{Si}{xi}]/[\ion{Si}{vi}] at $1.9\mu$m, have been thought to be ubiquitous of AGNs, concretely in the search for Intermediate Mass Black Holes (IMBHs) in the JWST era \citep{jr:cann2018ApJ...861..142C}.

In this vein, \citet{jr:molina2021ApJ...922..155M} performed a blind search for traces of [\ion{Fe}{x}]$\lambda$6374\AA~emission in SDSS galaxies and found up to 81 dwarfs with detectable [\ion{Fe}{x}] coronal line emission. \citet{jr:reefe2023ApJ...946L..38R} used MUSE IFS to report the detection of [\ion{Fe}{X}] in the brightest star-forming region of CGCG 007-025, with a flux value of $F_{\rm[\ion{Fe}{x}]} = (3.89\pm0.36) \times10^{-17}$~erg/s/cm$^2$/\AA, at $\geq10\sigma$ over the noise level. However, \citet{jr:herenz2023RNAAS...7...99H} point out that this detection is more likely to be a misidentification with the neighbouring \ion{Si}{ii}$\lambda$6371. This permitted line is part of a doublet (red companion at 6347\AA) and, with a much lower IP (16.35 eV), it is commonly found in the spectra of extragalactic SF regions \citep[e.g. ][]{jr:dominguezguzman2022MNRAS.517.4497D}.

\begin{figure}
    \centering
    \includegraphics[width=0.48\textwidth]{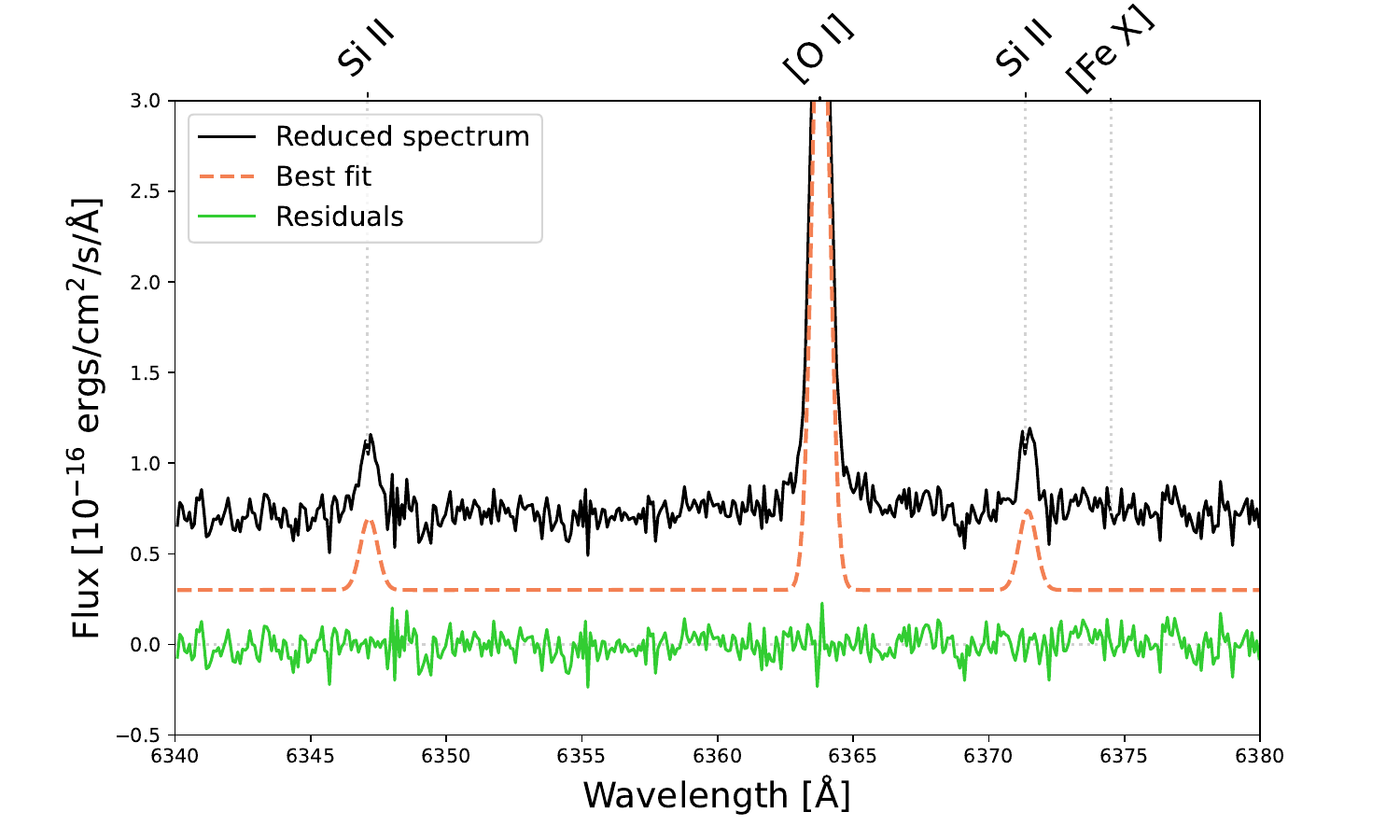}
    \caption{Spectral window from 6340 to 6380\AA. Marked with vertical lines are the rest-frame position of the emission lines known in this wavelength range, namely the [\ion{O}{i}]$\lambda$6363 line, \ion{Si}{ii}$\lambda\lambda$6347,6371 doublet and the [\ion{Fe}{x}]$\lambda$6374 line. The black line represents the reduced spectrum, with the best fit model displayed as a dashed-orange line and the residuals in solid-green. At the echelle resolution, the peak at 6371.44$\pm$0.03 \AA~matches undoubtedly the position of the \ion{Si}{ii} emission line. Note there is no clear evidence in the residuals of the necessity of another emission line around the [\ion{Fe}{x}] to account for emission from this element.}
    \label{fig:FeX}
\end{figure}

Our unique echelle spectrum provides sufficient resolution and signal to noise to revisit the case for a IMBH in CGCG 007-025: while the theoretical location of \ion{Si}{ii}$\lambda$6371 and [\ion{Fe}{x}]$\lambda$6374 are only separated by one MUSE spectral resolution element, there are 18 resolution elements between these two lines using our high-resolution (R$\sim$40,0000) echelle spectroscopy. In Figure \ref{fig:FeX}, the rest-frame window from 6340 to 6380 \AA\ is shown, where several emission lines are identified in the plot with dotted vertical lines. As one can see, the location of the emission line at 6371.44$\pm$0.03 \AA~coincides with the theoretical wavelength of \ion{Si}{ii}, with the line-detection at 6347.15$\pm$0.04\AA~also matching the rest-wavelength of the other component of the \ion{Si}{ii} doublet. 
Additionally, the intensity ratio of the two lines follows the theoretically-expected $\sim$1:1 ratio.

Based of this test, we conclude that the reported detection of coronal [\ion{Fe}{x}] by \citet{jr:reefe2023ApJ...946L..38R} is due to a misidentification with  \ion{Si}{ii}$\lambda$6371, as also suggested by \citet{jr:herenz2023RNAAS...7...99H}. It is true, however, that the possibility of [\ion{Fe}{X}] being blue-shifted to 6371\AA\ because of the presence of outflows cannot be ruled out. In this scenario, the velocity of the outflow would be 145 km/s. However, and once again due to our high spectral resolution (R$\sim$40,000), the minimum velocity shift to not be able to see a deblend between these lines would need to be only 14 km/s, which is still very unlikely. Finally, the absence of other coronal lines of similar IP, such as [\ion{Ne}{v}]$\lambda$3426 or [\ion{Fe}{vii}] to [\ion{Fe}{ix}] in the optical range, also supports our hypothesis.

\subsection{The S/O, Ar/O and Ne/O abundances}

O, S, Ne and Ar are known as $\alpha$-elements. This elements are synthesised in massive stars by $\alpha$-particle capture and released to the ISM by core-collapse supernovae. The abundances of these elements should change in accordance, implying that the abundance ratios of S, Ne and Ar with respect to O should be constant across galaxies \citep[e.g., ][]{jr:Izotov2006A&A...448..955I,jr:Guseva2011A&A...529A.149G,jr:Miranda-Perez2023ApJ...952...76M} and \ion{H}{ii} regions \citep[e.g., ][]{jr:Croxall2015ApJ...808...42C,jr:arellano-cordova2020MNRAS.496.1051A,jr:Rogers2022ApJ...939...44R,jr:DiazZamora2022MNRAS.511.4377D}. Using our derived abundances, we can calculate the abundance ratios with respect to the oxygen values to be:

\begin{equation*}
    \begin{array}{r}
        \text{log(S/O)} = -1.66 \pm 0.06\\
        \text{log(Ar/O)} = -2.35 \pm 0.04\\
        \text{log(Ne/O)} = -0.70 \pm 0.05\\
    \end{array}
\end{equation*}

Using low resolution spectroscopy from SDSS, \citet{jr:Izotov2006A&A...448..955I} estimated the abundance of S, Ne and Ar in the same region of study in this paper. They derived log(S/O) = -1.68$\pm$0.04, log(Ar/O) = -2.40$\pm$0.04 and log(Ne/O) = -0.85 $\pm$0.04. Our abundances for S and Ar are in very good agreement, but the Ne/O seems to be slightly higher than in previous studies for this object. The CLASSY collaboration has also estimated the S, Ne and Ar relative abundances using the spectrum from MODS/LBT \citep{jr:arellano-cordova2024ApJ...968...98A}. The values they recover are log(Ne/O) =  -0.70$\pm$0.01, log(S/O) = -1.66$\pm$0.02, y log(Ar/O) = -2.29$\pm$0.02, also in excellent agreement with our estimations. 
However, in \citet{jr:dominguezguzman2022MNRAS.517.4497D} the values of 8 \ion{H}{ii} regions in the LMC and SMC for the abundances ratios were reported, as well as compared with the solar values from \citet{jr:Lodders2019arXiv191200844L}. For the sulphur, the log(S/O) and log(Ar/O) values, for both the Magellanic Clouds (MCs) and the MW, are also compatible with our abundances from the echelle data: log(S/O)$_\text{MCs}$ = -1.66$\pm$0.02 and log(S/O)$_\text{MW}$=-1.58$\pm$0.08, log(Ar/O)$_\text{MCs}$=-2.32$\pm$0.05 and log(Ar/O)$_\text{MW}$=-2.23$\pm$0.12. For the Neon-to-Oxygen ratio, the reported values are log(Ne/O)$_\text{MCs}$=-0.57$\pm$0.02 and log(Ne/O)$_\text{MW}$=-0.58$\pm$0.12. 
Although our log(Ne/O) estimation is still lower than the MCs and MW, it is compatible within errors with this nearby high-resolution spectroscopy study, as well as with other SFGs with intermediate resolution spectra \citep{jr:berg2019ApJ...874...93B}.  

 Low Ne/O values are also reported in H II regions of disc galaxies. Such lower Ne/O could be related to the ionization correction factors \citep[e.g.,][]{jr:amayo2021MNRAS.505.2361A,jr:Izotov2006A&A...448..955I} that might fail to reproduce the total abundance of Ne \citep[e.g.,][]{jr:croxall2016ApJ...830....4C}. However, the low log(Ne/O) value for CGCG 00-025 is compatible with the errors of another nearby high-resolution spectroscopy study, as well as with other SFGs with intermediate resolution spectra \citep[e.g.,][]{jr:berg2019ApJ...874...93B,jr:arellano-cordova2024ApJ...968...98A}. In addition,  \citet{jr:isobe2023ApJ...959..100I} also reported low Ne/O ratios in a sample of high redshift galaxies. These low values might be explained by considering chemical evolution models of stars with stellar mass $\gtrsim$30 $M_{\odot}$ \citep[see also][]{jr:watanabe2024ApJ...962...50W}.

\subsection{The Fe/O abundance ratio}

Metals are formed in stars and are ejected into the ISM by strong stellar winds and supernova explosions. More than 90\% of the baryons are found in the gas phase at z$>$2 \citep{jr:PerouxHowk2020ARA&A..58..363P}. These metals in the ISM can be incorporated in to the next generation of stars. While substantial amount of metals are found in neutral gas, large fractions of these metals are instead locked into dust grains, an effect called dust depletion \citep{jr:SavageSembach1996ApJ...470..893S,jr:DeCia2016A&A...596A..97D,jr:Roman-Duval2021ApJ...910...95R}

The high depletion factors found for Fe in the interstellar medium(ISM), down to [Fe/H]=-2.3 \citep{jr:SavageSembach1996ApJ...470..893S}, and the relatively high cosmic abundance of this element imply that Fe is a very important contributor to the mass of refractory dust grains \citep{jr:Sofia1994ApJ...430..650S}.
Several studies have found the Fe/O ratio is typically well below solar \citep[][among others]{jr:rodriguez2005ApJ...626..900R,jr:Izotov2006A&A...448..955I}, with a decrease of the Fe/O abundance ratio with increasing O/H --  which implies that depletion of Fe increases with increasing metallicity.

As part of the sample, \citet{jr:Izotov2006A&A...448..955I} estimate a Fe/O for the object of study of log(Fe/O) = -1.93$\pm$0.14. Using \citet{jr:Izotov2006A&A...448..955I} definition for the ICF($Fe^{2+}$) and our $Fe^{2+}/H^+$ derived from [\ion{Fe}{iii}]$\lambda$4658\AA, we estimate an iron abundance relative to oxygen of log(Fe/O) = -1.97$\pm$0.04. Moreover, \citet{jr:rodriguez2005ApJ...626..900R} found that galaxies with a metallicity close to the metallicity of the SMC have a depletion factor [Fe/O] \footnote{defined as [Fe/O] = log(Fe/O)-log(Fe/O)$_\odot$, with log(Fe/O)$_\odot$=-1.28 \citep{jr:Lodders2019arXiv191200844L}} in the range -0.5 to -1.1. When compared to the solar value, the depletion factor of CGCG 007-025 is [Fe/O] = -0.69, well in the range expected for this metallicity.

\subsection{The origin of the HeII emission}

The ionisation of \ion{He}{ii} requires environments capable of generating ionising photons with energies above 54.4 eV. Observationally, the detection of \ion{He}{ii}$\lambda$4686 in star-forming galaxies suggests that processes related to star formation are able to create such environments. However, the exact nature of the sources responsible for this nebular line is still unclear.

Among the candidates, Wolf-Rayet (WR) stars used to stand as the primary contender \citep{jr:Schaerer1996ApJ...467L..17S,jr:Senchyna2022}. During this phase, large quantities of photons are emitted with sufficient energies to ionise \ion{He}{ii}. Although single stellar population (SSP) models are able to explain the observed \ion{He}{ii}$\lambda$4686/H$\beta$ intensity ratios in some star-forming galaxies \citep{jr:Plat2019MNRAS.490..978P}, these models predict a maximum value of \ion{He}{ii}$\lambda$4686/H$\beta$ = 0.01 during the WR phase for metallicities Z $\geq$0.004, with this ratio dropping further at lower metallicities. Yet, the observed values of the \ion{He}{ii}$\lambda$4686/H$\beta$ ratio in metal-poor galaxies often are found to be 0.01-0.1 \citep{jr:Kehrig2015ApJ...801L..28K,jr:Kehrig2018MNRAS.480.1081K,jr:schaerer2019A&A...622L..10S}. Models incorporating binary stars help to alleviate the problem to some extent, but explaining the presence of \ion{He}{ii} in high H$\beta$ equivalent width (EW) systems remains a challenge \citep{jr:Plat2019MNRAS.490..978P}. \textcolor{corrections}{More recently, hot stars produced by mass transfer in binary systems have been shown to boost the \ion{He}{ii}/H$\beta$ ratios at high H$\beta$ equivalent widths \citep{jr:lecroq2024MNRAS.527.9480L}. Still, for all binary models the predicted \ion{He}{ii}/H$\beta$ ratios are lower than the measured value} \textcolor{corrections3}{of the medium components of \ion{He}{ii} and H$\beta$} in our ultra-high resolution spectrum.

\citet{jr:schaerer2019A&A...622L..10S} found that the observed \ion{He}{ii}$\lambda$4686/H$\beta$ ratio in metal-poor galaxies can be explained if the bulk of the \ion{He}{ii} ionizing photons is emitted by HMXBs, whose numbers are found to increase with decreasing metallicity, and where the ratio between Q(He+)\footnote{defined as the number of ionising photons above 54.4 eV.} to X-ray luminosity appears to be constant. However, \citet{jr:Plat2019MNRAS.490..978P} warned that this process is not efficient at EW(H$\beta$) $>$ 200 \AA, as these systems are too young to form compact objects (such as neutron stars and stellar mass black holes) necessary for the existence of HMXBs. %Our region of study has EW(H$\beta$)=290\AA~(from \citetalias{jr:mgve2023}). 
In a previous study, \citet{jr:Senchyna2020MNRAS.494..941S} explored this possibility using Keck/ESI data of CGCG 007-025. With a ratio \ion{He}{ii}$\lambda$4686/H$\beta$ = (1.30$\pm$0.06)$\times 10^{-2}$ and
an X-ray production efficiency of $1.47 \times 10^{40}$ erg s$^{-1}$, they concluded that HMXBs do not produce enough ionising flux to fully explain the observed \ion{He}{ii} emission in this SF region.

\begin{figure*}
    \centering
    \includegraphics[width=0.98\textwidth]{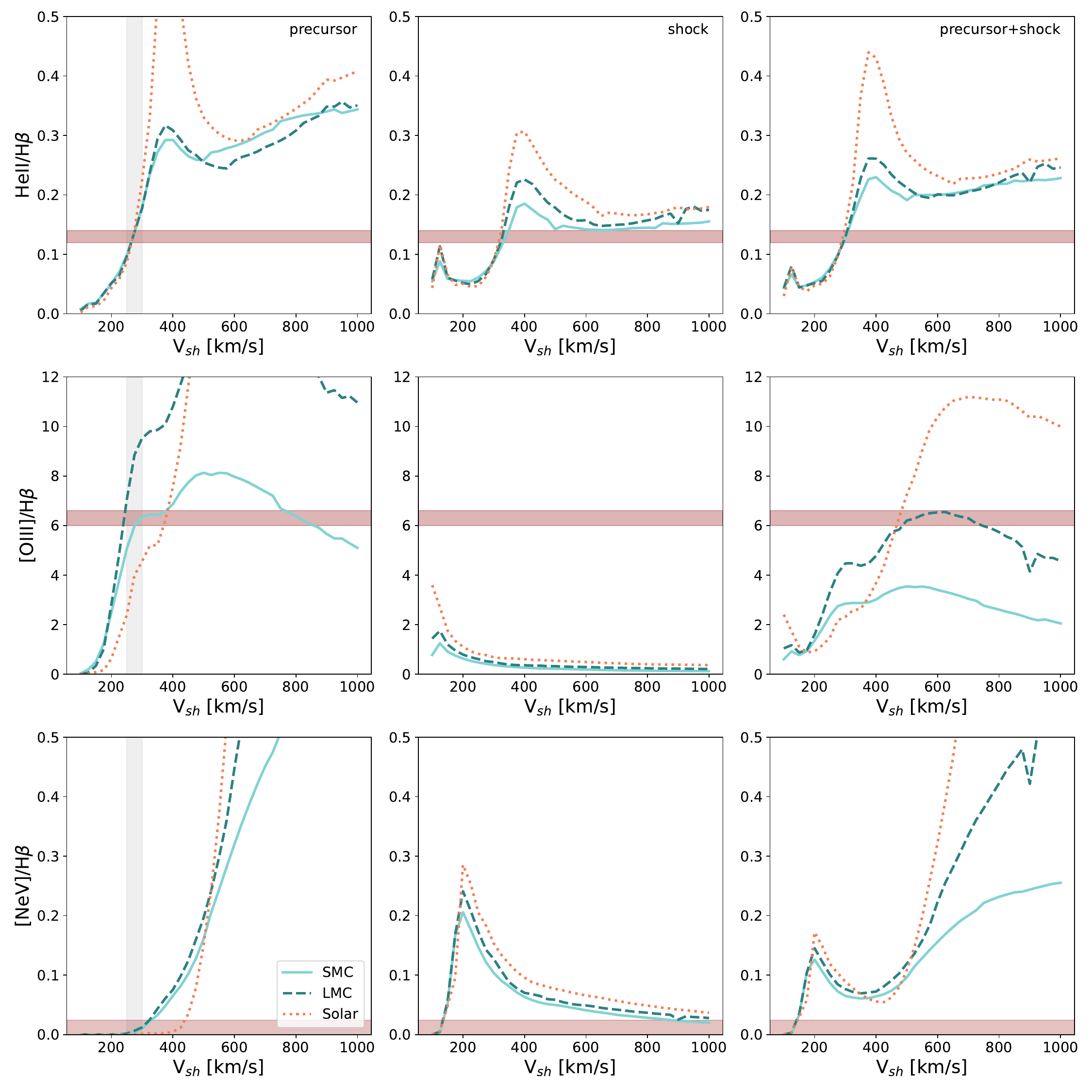}
    \caption{Production of \ion{He}{ii}$\lambda$4686/H$\beta$ (top), [\ion{O}{iii}]$\lambda$5007/H$\beta$ (middle) and [\ion{Ne}{v}]$\lambda$3426/H$\beta$ (bottom) depending on the velocity of the shock as modelled by \citet{jr:allen2008ApJS..178...20A}. The left column correspond to precursor only models, the central column to shock only models and the right column to the addition of precursor and shock. The solid-cyan line corresponds to the SMC metallicity models, the dashed-teal to the LMC, and the dotted-orange to solar abundances. Our flux measurements on \ion{He}{ii}$\lambda$4686/H$\beta$ and [\ion{O}{iii}]$\lambda$5007/H$\beta$ are highlighted with the brown horizontal band, and an 3$\sigma$ upper limit on the detection of [\ion{Ne}{v}]$\lambda$3426/H$\beta$ is marked with a light brown band shaded with diagonal lines. The light grey vertical band indicates the shock velocities that explain the observed line ratios, assuming the precursor-only SMC-metallicity models are the most appropriate in our case.}
    \label{fig:shocks}
\end{figure*}

One of the most promising explanations for the hard radiation and the high-ionisation emission lines is the presence of fast radiative shocks. 
In fast shocks, the ionising radiation produced by the cooling of hot gas behind the shock front generates a strong radiation field consisting of extreme ultraviolet and soft X-ray photons, resulting in significant photoionisation effects. For velocities larger than $v_s\approx170$ km s$^{-1}$, the ionisation front surpasses and detaches from the shock front \citep{jr:allen2008ApJS..178...20A}. This detachment leads to the expansion of a precursor \ion{H}{ii} region ahead of the shock. Moreover, at higher shock velocities, the emission from the photoionised precursor may overshadow the optical emission from the shock. In order to explore this scenario, we use the models from \citet{jr:allen2008ApJS..178...20A} \textcolor{corrections}{--with a conservative value of 0.5$\mu$G for the magnetic field--} to compute 
the production of \ion{He}{ii}$\lambda$4686/H$\beta$, [\ion{Ne}{v}]$\lambda$3426/H$\beta$ and [\ion{O}{iii}]$\lambda$5007/H$\beta$ for the pre-shock region, shock, and precursor+shock depending on the velocity of the shock for three different metallicities (solar abundances, LMC and SMC) -- as shown in Figure \ref{fig:shocks}. 
The different emission line ratios are displayed in rows, while the components of the shock are displayed in columns. \textcolor{corrections}{We remark that the emission of \ion{He}{ii}$\lambda$4686 from our region of study is solely described by a gaussian component of $\sigma\approx$ 35 km s$^{-1}$, hereafter intermediate component. As such, we only use the line ratios compatible with this kinematical component for our comparison with the models. 
Our derived value of \ion{He}{ii}$\lambda$4686/H$\beta$ = 0.13 $\pm$ 0.01 --highlighted in brown-- can be well explained by all the considered models: precursor, shock and precursor+shock at any metallicity. However, when looking at the [\ion{Ne}{v}]$\lambda$3426/H$\beta$ and [\ion{O}{iii}]$\lambda$5007/H$\beta$ graphs, our measured values (i.e., [\ion{Ne}{v}]$\lambda$3426 non detection and [\ion{O}{iii}]$\lambda$5007/H$\beta$=6.3$\pm$0.3) are only reproduced by the precursor model (left-column panels), concretely at low metallicities (SMC).} 

\textcolor{corrections}{As a conclusion, the \ion{He}{ii} emission in this SF region is likely arising from the ionisation front preceding a shock. Our measurements associated with this intermediate component can additionally constrain the velocity of the shock front to be in the range $v_{sh}$=250-300 km s$^{-1}$. At these velocities, the expected temperature of the post-shock gas is around $10^5-10^6$ K \citep[equation 36.38 from][]{2011piim.book.....D}. 
Moreover, for moderately fast shocks ($v_s$ > 200 km s$^{-1}$), the expected line broadening should be comparable to or above the shock velocity (i.e., in the range 300-1000 km s$^{-1}$). Components of such line widths will not be visible at this spectral resolution, falling within the continuum noise at this depth} \textcolor{corrections3}{(see Figure \ref{fig:hbeta} for an illustrated example)}. Conversely, the observed $\sim$35 km/s components \ion{He}{ii}, [\ion{O}{iii}] or H$\beta$ can alternatively be attributed to the cooler and less disturbed gas in the precursor (also known as the photoionisation front), and with expected velocity dispersion in the range \textcolor{corrections3}{$\lesssim$50-100 km s$^{-1}$} \citep{jr:izotov2012MNRAS.427.1229I}.

\textcolor{corrections}{These results highlight the need for high resolution spectroscopy ($\leq$10 km s$^{-1}$) to disentangle the ionisation mechanisms within extragalactic \ion{H}{ii} regions, where different kinematic components could have different origins. For example, the lack of ability to resolve the components presented in this study would result in a ratio \ion{He}{ii}$\lambda$4686/H$\beta_\text{all}$ = 0.012 $\pm$ 0.003, in agreement with previous measurements at lower resolutions \citep{jr:Senchyna2020MNRAS.494..941S} but inconsistent with the aforementioned models.}

\section{Conclusions}
\label{sec:Conclusions}

In this paper, we use echelle spectroscopy to derive the chemical properties of the brightest SF region in the metal-poor dwarf galaxy CGCG 007-025. To the best of our knowledge, this is one of the highest resolution spectrum ever studied not only for this galaxy, but also for comparable metal-poor objects. Our main results are summarised below:

\begin{itemize}
    \item The exceptionally high resolution (R$\sim$40,000) of the echelle spectrum allows us to detect and resolve up to 80 unique emission lines. Most of the low-intensity lines are well modelled with a single Gaussian profile with $\sigma_{\rm narrow}=14$ km/s, while brighter emission lines require an intermediate component with $\sigma_{\rm medium}=33$km/s. The brightest emission lines are kinematically complex, requiring up to four different components: three to describe the core and wings of the main peak, and a secondary (fainter) narrow component to describe a kinematically-decoupled (redshifted) emission along the same line of sight.  
    \item We derive the chemical properties of the SF region using a two-region ionisation model and the simultaneous fit to 30 emission lines. The electron temperatures $T_{low}$ and $T_{high}$ are in good agreement with previous studies. The gas-phase metallicity of the region is $12+\log(O/H)=7.77\pm 0.03$, placing the object in the metal-poor regime. The sulphur abundance we report is $\log(S/H)=6.1\pm0.1$. Both of these values are in good agreement with previous studies as well.
    \item When looking at the Metals-to-Oxygen ratios for the $\alpha$-elements, i.e. the S/O, the Ar/O and the Ne/O, our derived values are in excellent agreement with previous studies of this object using lower spectral resolution. Moreover, these ratios are also in good agreement with other echelle studies done in \ion{H}{ii}-regions of the Milky Way and Magellanic Clouds.  Additionally, the log(Fe/O) ratio measured for this SF region is in agreement with previous studies, with the [Fe/O] value falling in the range expected at this metallicity.
    \item We detect and resolve \ion{He}{ii}$\lambda$4686 emission. The line  presents unusual properties, including clear asymmetry towards high velocities, and a velocity dispersion that is well reproduced by a single, wide kinematic component ($\sigma\sim$\,33 km/s; cf. $\sigma_{narrow}\sim$\,14 km/s and $\sigma_{medium}\sim$\,37 km/s for most of the other lines). %The exceptionally high resolution (R$\sim$40,000) of this echelle spectrum allows us to not only detect but also to resolve \ion{He}{ii}$\lambda$4686. This line is well modelled with a single Gaussian profile, whose velocity dispersion is higher than the narrow component of the other lines and that actually matches the velocity dispersion of the intermediate. 
    By comparing the \ion{He}{ii}/H$\beta$ and [\ion{O}{iii}]/H$\beta$ ratios of the intermediate components, together with the absence of other high ionisation lines such us [\ion{Ne}{v}], the presence of fast radiative shocks with a velocity in the range of $v_{\rm sh} = 250-300$ km/s are revealed.
\end{itemize}

Their proximity and characteristics make nearby metal-poor starburst dwarfs ideal laboratories to probe, with unprecedented spatial and spectral resolution, the physics of high-density star formation akin to what is found in the high-z Universe. 
The echelle spectrum presented in this paper has proved an essential tool to disentangle the superposition of different ionisation scenarios as well as to constrain with superb detail the chemical structure of this metal-poor star-forming region. %evidences the power of echelle spectroscopy to provide detailed chemodynamical analysis of SF regions in the Local Volume. Echelle spectrocopy has proved to be an essential tool to disentangle superposition of ionisation scenarios in SF galaxies.

\section*{Acknowledgements}

The authors thank Carolina Kehrig, Antonio Arroyo Polonio, Daniel Schaerer, Cesar Esteban and Matthew J. Hayes for the fruitful discussions on the origin of the \ion{He}{ii} emission.
M.G.V.E acknowledges the support of the UK Science and Technology Facilities Council. 
V.F. acknowledges the support by the Eric and Wendy Schmidt AI in Science Postdoctoral Fellowship, a Schmidt Futures program, at the Michigan Institute for Data Science, University of Michigan. RA acknowledges financial support from the State Agency for Research of the Spanish MCIU through ‘Center of Excellence Severo Ochoa’ award to the IAA-CSIC (SEV-2017-0709) and CEX2021-001131-S funded by MCIN/AEI/10.13039/501100011033, and from projects PID2023-147386NB-I00 "XTREM" and PID2022- 136598NB-C32 “Estallidos8”. The work of KB is supported by NOIRLab, which is managed by the Association of Universities for Research in Astronomy (AURA) under a cooperative agreement with the U.S. National Science Foundation.

\textit{Software}: this work made an extensive use of \textsc{Python}, and more specifically of \textsc{lmfit} \citep{pr:lmfit}, \textsc{numpy} \citep{pr:numpy}, \textsc{astropy} \citep{pr:astropy}, \textsc{scipy}, \textsc{matplotlib} \citep{pr:matplotlib} and \textsc{lineid\_plot} \citep{pr:lineid_plot}.

%%%%%%%%%%%%%%%%%%%%%%%%%%%%%%%%%%%%%%%%%%%%%%%%%%
\section*{Data Availability}

\textcolor{corrections}{This paper includes data gathered with the 6.5 meter Magellan Telescopes located at Las Campanas Observatory (LCO), Chile. The data underlying this work are available in the article as tables.} 
%The inclusion of a Data Availability Statement is a requirement for articles published in MNRAS. Data Availability Statements provide a standardised format for readers to understand the availability of data underlying the research results described in the article. The statement may refer to original data generated in the course of the study or to third-party data analysed in the article. The statement should describe and provide means of access, where possible, by linking to the data or providing the required accession numbers for the relevant databases or DOIs.

%%%%%%%%%%%%%%%%%%%% REFERENCES %%%%%%%%%%%%%%%%%%

% The best way to enter references is to use BibTeX:

\bibliographystyle{mnras}
\bibliography{mnras} % if your bibtex file is called example.bib

% Alternatively you could enter them by hand, like this:
% This method is tedious and prone to error if you have lots of references
%\begin{thebibliography}{99}
%\bibitem[\protect\citeauthoryear{Author}{2012}]{Author2012}
%Author A.~N., 2013, Journal of Improbable Astronomy, 1, 1
%\bibitem[\protect\citeauthoryear{Others}{2013}]{Others2013}
%Others S., 2012, Journal of Interesting Stuff, 17, 198
%\end{thebibliography}

%%%%%%%%%%%%%%%%%%%%%%%%%%%%%%%%%%%%%%%%%%%%%%%%%%

%%%%%%%%%%%%%%%%% APPENDICES %%%%%%%%%%%%%%%%%%%%%

\appendix

\section{Figures}

\begin{figure*}
  \sbox0{\begin{tabular}{@{}cc@{}}
    \includegraphics[width=0.78\textheight]{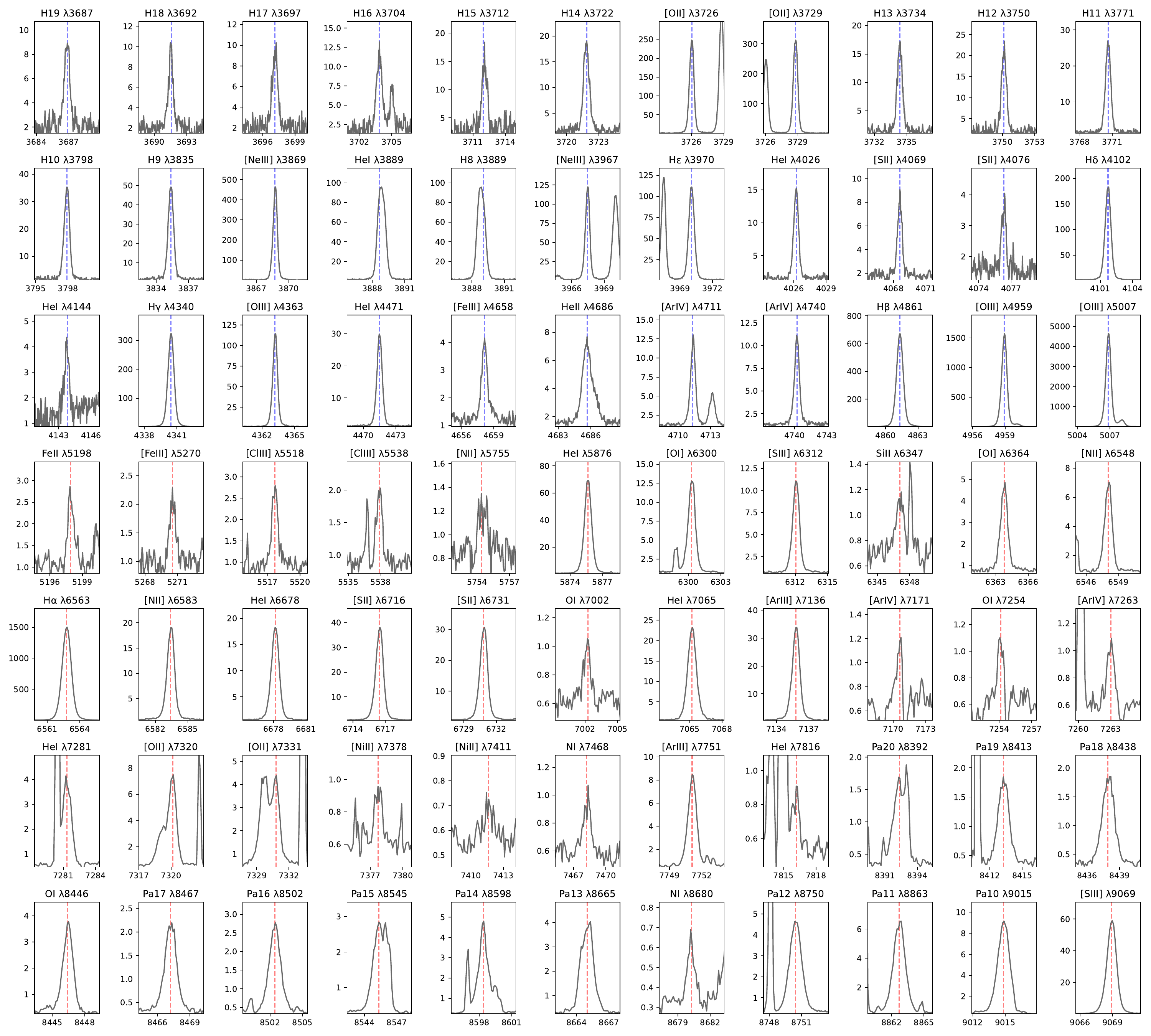}
    \label{fig:emissionLines}
  \end{tabular}}% measure width
  \rotatebox{90}{
  \begin{minipage}[c][\textwidth][c]{\wd0}
    \usebox0
    \caption{Compilation of emission lines detected in the echelle spectrum with AON$>$5. Lines detected in the blue (red) arm are marked with a blue(red)-dashed line. Flux units (y-axis) are $10^{-16}$ erg s$^{-1}$ cm$^{-2}$ \AA$^{-1}$. Wavelength  units (x-axis) are \AA, in rest-frame.}
  \end{minipage}
}
\end{figure*}

\begin{figure}
    \centering
    \includegraphics[width=0.5\textwidth]{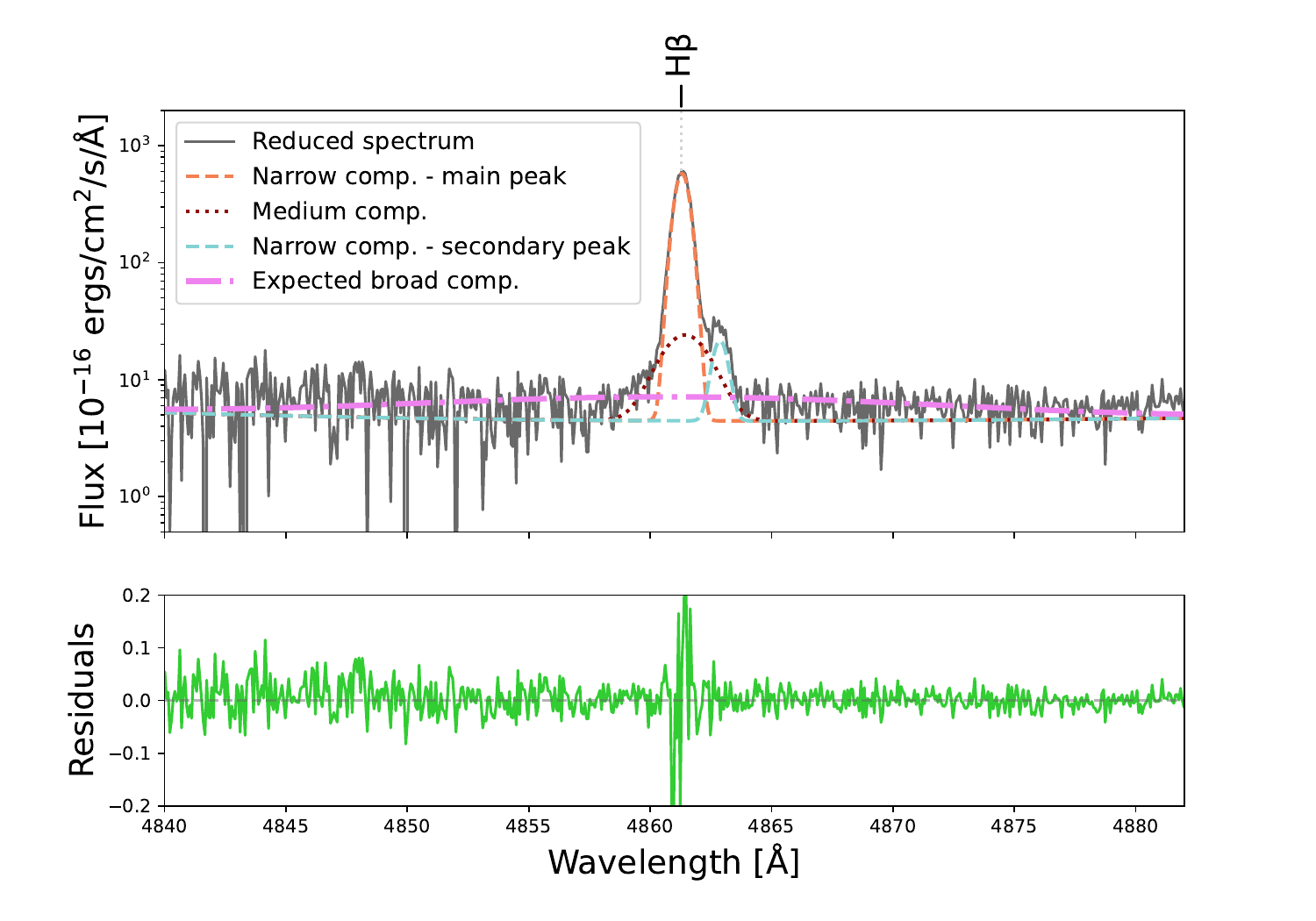}
    \caption{\textcolor{corrections3}{Zoom in into the H$\beta$ spectral region in the red arm. The reduced spectrum is shown in grey, with the different components required to reproduce the line profile in dashed-orange (main narrow component), dashed-blue (secondary narrow component) and dotted-maroon (medium component). The residuals are displayed in green in the bottom panel. The pink-dotted-dashed line represents the expected emission from the shock assuming a width of $\sim$650 km/s and the flux prediction from MAPPINGS. This exercise illustrates the lack of ability of our dataset in resolving such component. }}
    \label{fig:hbeta}
\end{figure}

\section{Tables}

\begin{table}
    \centering
    \begin{tabular}{ccccc}
        $\lambda_0$ & Ion               & Flux & $\lambda_{\text{obs}}$ [\AA]   & $\sigma$ [\AA] \\ \hline \hline
        3686 & H19                      &   3.24$\pm$ 0.33 & 3686.82$\pm$0.03   & 0.32$\pm$0.03 \\
        3691 & H18                      &    3.04$\pm$0.26 & 3691.49$\pm$0.03   & 0.29$\pm$0.03 \\
        3697 & H17                      &    3.70$\pm$0.27 & 3697.17$\pm$0.02   & 0.28$\pm$0.02 \\
        3703 & H16                      &    4.21$\pm$0.41 & 3703.83$\pm$0.03   & 0.26$\pm$0.03 \\
        3711 & H15                      &    3.49$\pm$0.63 & 3711.92$\pm$0.05   & 0.22$\pm$0.05 \\
        3721 & H14                      &    8.13$\pm$0.26 & 3721.80$\pm$0.01   & 0.28$\pm$0.01 \\
        3726 & [OII]$_{\rm n}$          &   82.16$\pm$1.06 & 3726.01$\pm$0.01   & 0.20$\pm$0.01 \\
        3726 & [OII]$_{\rm m}$          &   10.10$\pm$0.72 & 3726.05$\pm$0.02   & 0.61$\pm$0.04 \\
        3728 & [OII]$_{\rm n}$          &  106.2$\pm$ 1.47 & 3728.76$\pm$0.01   & 0.20$\pm$0.01 \\
        3728 & [OII]$_{\rm m}$          &  10.81$\pm$ 1.13 & 3728.74$\pm$0.02   & 0.51$\pm$0.03 \\
        3734 & H13                      &   6.39$\pm$ 0.18 & 3734.34$\pm$0.01   & 0.25$\pm$0.01 \\
        3750 & H12                      &   7.70$\pm$ 0.64 & 3750.09$\pm$0.02   & 0.21$\pm$0.02 \\
        3770 & H11                      &  10.68$\pm$ 0.17 & 3770.60$\pm$0.01   & 0.24$\pm$0.01 \\
        3797 & H10                      &  14.85$\pm$ 0.29 & 3797.87$\pm$0.01   & 0.25$\pm$0.01 \\
        3835 & H9                       &  20.56$\pm$ 0.51 & 3835.36$\pm$0.01   & 0.25$\pm$0.01 \\
        3868 & [NeIII]$_{\rm n}$        & 155.64$\pm$ 5.97 & 3868.73$\pm$0.01   & 0.21$\pm$0.01 \\
        3868 & [NeIII]$_{\rm m}$        &   6.18$\pm$ 3.06 & 3869.26$\pm$0.30   & 0.49$\pm$0.03 \\
        3888 & HeI                      &  25.13$\pm$ 1.30 & 3888.64$\pm$0.01   & 0.21$\pm$0.01 \\
        3889 & H8$_{\rm n}$             &  22.67$\pm$ 1.21 & 3889.06$\pm$0.01   & 0.21$\pm$0.01 \\
        3889 & H8$_{\rm m}$             &  13.02$\pm$ 2.20 & 3888.87$\pm$0.02   & 0.49$\pm$0.03 \\
        3967 & [NeIII]$_{\rm n}$        &  34.06$\pm$ 2.31 & 3967.46$\pm$0.01   & 0.18$\pm$0.01 \\
        3967 & [NeIII]$_{\rm m}$        &  12.49$\pm$ 2.22 & 3967.40$\pm$0.02   & 0.36$\pm$0.03 \\
        3970 & H$\varepsilon_{\rm n}$   &  48.89$\pm$ 0.87 & 3970.05$\pm$0.01   & 0.25$\pm$0.01 \\
        3970 & H$\varepsilon_{\rm m}$   &   5.10$\pm$ 0.68 & 3969.90$\pm$0.05   & 0.72$\pm$0.08 \\
        4026 & HeI                      &   5.55$\pm$ 0.12 & 4026.19$\pm$0.01   & 0.22$\pm$0.01 \\
        4068 & [SII]                    &   2.44$\pm$ 0.10 & 4068.59$\pm$0.01   & 0.21$\pm$0.01 \\
        4076 & [SII]                    &   0.75$\pm$ 0.08 & 4076.32$\pm$0.03   & 0.23$\pm$0.03 \\
        4101 & H$\delta_{\rm n}$        &  81.25$\pm$ 1.87 & 4101.71$\pm$0.01   & 0.26$\pm$0.01 \\
        4101 & H$\delta_{\rm m}$        &  12.06$\pm$ 1.49 & 4101.76$\pm$0.03   & 0.64$\pm$0.04 \\
        4143 & HeI                      &   1.78$\pm$ 0.27 & 4143.82$\pm$0.08   & 0.43$\pm$0.08 \\
        4340 & H$\gamma_{\rm n}$        & 164.00$\pm$ 2.22 & 4340.44$\pm$0.01   & 0.27$\pm$0.01 \\
        4340 & H$\gamma_{\rm m}$        &  21.49$\pm$ 1.35 & 4340.47$\pm$0.01   & 0.68$\pm$0.02 \\
        4363 & [OIII]$_{\rm n}$         &  40.70$\pm$ 0.88 & 4363.19$\pm$0.01   & 0.21$\pm$0.01 \\
        4363 & [OIII]$_{\rm m}$         &   9.16$\pm$ 0.72 & 4363.18$\pm$0.01   & 0.51$\pm$0.02 \\
        4471 & HeI$_{\rm n}$            &  12.16$\pm$ 0.60 & 4471.49$\pm$0.01   & 0.23$\pm$0.01 \\
        4471 & HeI$_{\rm m}$            &   2.33$\pm$ 0.56 & 4471.44$\pm$0.04   & 0.53$\pm$0.07 \\
        4658 & [FeIII]                  &   1.37$\pm$ 0.07 & 4658.13$\pm$0.01   & 0.27$\pm$0.01 \\
        4686 & HeII                     &   5.44$\pm$ 0.57 & 4685.71$\pm$0.06   & 0.51$\pm$0.06 \\
        4711 & [ArIV]$_{\rm n}$         &   4.75$\pm$ 0.11 & 4711.36$\pm$0.01   & 0.23$\pm$0.01 \\
        4711 & [ArIV]$_{\rm m}$         &   1.15$\pm$ 0.18 & 4711.42$\pm$0.25   & 1.50$\pm$0.04 \\
        4713 & HeI                      &   1.93$\pm$ 0.09 & 4713.16$\pm$0.01   & 0.24$\pm$0.01 \\
        4740 & [ArIV]$_{\rm n}$         &   4.23$\pm$ 0.11 & 4740.20$\pm$0.01   & 0.23$\pm$0.01 \\
        4740 & [ArIV]$_{\rm m}$         &   1.75$\pm$ 0.16 & 4740.53$\pm$0.14   & 1.50$\pm$0.04 \\
        4861 & H$\beta_{\rm n}$         & 397.05$\pm$ 4.43 & 4861.30$\pm$0.01   & 0.31$\pm$0.01 \\
        4861 & H$\beta_{\rm m}$         &  40.46$\pm$ 2.64 & 4861.36$\pm$0.02   & 0.80$\pm$0.02 \\
        4861 & H$\beta^{\star}_{\rm n}$ &   1.75$\pm$ 0.51 & 4862.45$\pm$0.06   & 0.31$\pm$0.01 \\
        4958 & [OIII]$_{\rm n}$         &  706.37$\pm$16.60 & 4958.90$\pm$0.01 & 0.26$\pm$0.01 \\
        4958 & [OIII]$_{\rm m}$         &  84.35$\pm$ 3.91 & 4958.98$\pm$0.02   & 0.68$\pm$0.01 \\
        4958 & [OIII]$_{\rm b}$         &  10.60$\pm$ 0.34 & 4960.73$\pm$0.17   & 5.8 $\pm$0.2  \\
        4958 & [OIII]$^{\star}_{\rm n}$ &  17.24$\pm$ 1.04 & 4960.17$\pm$0.01   & 0.26$\pm$0.01 \\
        5006 & [OIII]$_{\rm n}$         & 2084.54$\pm$42.46 & 5006.83$\pm$0.01 & 0.25$\pm$0.01 \\
        5006 & [OIII]$_{\rm m}$         & 253.04$\pm$11.73 & 5006.91$\pm$0.02   & 0.68$\pm$0.01 \\
        5006 & [OIII]$_{\rm b}$         &  31.80$\pm$ 1.01 & 5008.68$\pm$0.17   & 5.8 $\pm$0.2  \\
        5006 & [OIII]$^{\star}_{\rm n}$ &   52.07$\pm$ 1.46 & 5008.45$\pm$0.01 & 0.25$\pm$0.01 \\
        5015 & HeI                      &    7.26$\pm$ 0.24 & 5015.69$\pm$0.01 & 0.28$\pm$0.01 \\

    \end{tabular}
    \caption{Best fit parameters of the model lines in the blue arm. Column (1): theoric. Column (2): ion. Column (3): observed flux. Column (4): rest-frame wavelength [\AA]. Column (5): measured width [\AA]. The sub-indexes n, m and b correspond to the narrow, intermediate and broad components respectively. Lines marked with $\star$ denote the secondary red-shifted peak described in Section \ref{sec:multipleLines}.}
    \label{tab:lineFitsBLUE}
\end{table}

\begin{table}
    \centering
    \begin{tabular}{ccccc}
        $\lambda_0$ & Ion      & Flux & $\lambda_{\text{obs}}$ [\AA] & $\sigma$ [\AA] \\  \hline \hline
        5197 & FeII                   &    0.71$\pm$ 0.23 & 5197.83$\pm$0.09 & 0.25$\pm$0.08 \\
        5270 & [FeIII]                &    0.79$\pm$ 0.09 & 5270.51$\pm$0.04 & 0.31$\pm$0.04 \\
        5517 & [ClIII]                &    1.06$\pm$ 0.07 & 5517.65$\pm$0.02 & 0.27$\pm$0.02 \\
        5537 & [ClIII]                &    0.75$\pm$ 0.08 & 5537.89$\pm$0.03 & 0.26$\pm$0.03 \\
        5754 & [NII]                  &    0.35$\pm$ 0.08 & 5754.68$\pm$0.08 & 0.33$\pm$0.08 \\
        5875 & HeI$_{\rm n}$          &   43.64$\pm$ 1.21 & 5875.65$\pm$0.01 & 0.31$\pm$0.01 \\
        5875 & HeI$_{\rm m}$          &    9.04$\pm$ 1.06 & 5875.64$\pm$0.02 & 0.65$\pm$0.03 \\
        6300 & [OI]$_{\rm n}$         &    8.01$\pm$ 0.32 & 6300.28$\pm$0.01 & 0.30$\pm$0.01 \\
        6300 & [OI]$_{\rm m}$         &    1.92$\pm$ 0.28 & 6300.06$\pm$0.10 & 0.86$\pm$0.10 \\
        6312 & [SIII]                 &    7.69$\pm$ 0.14 & 6312.06$\pm$0.01 & 0.32$\pm$0.01 \\
        6347 & SiII                   &    0.28$\pm$ 0.04 & 6347.10$\pm$0.02 & 0.25$\pm$0.04 \\
        6363 & [OI]                   &    3.10$\pm$ 0.09 & 6363.75$\pm$0.01 & 0.35$\pm$0.01 \\
        6371 & SiII                   &    0.34$\pm$ 0.06 & 6371.34$\pm$0.06 & 0.31$\pm$0.06 \\
        6548 & [NII]$_{\rm n}$        &    4.60$\pm$ 0.06 & 6548.04$\pm$0.01 & 0.33$\pm$0.01 \\
        6548 & [NII]$_{\rm m}$        &    0.56$\pm$ 0.04 & 6548.09$\pm$0.09 & 1.50$\pm$0.13 \\
        6562 & H$\alpha_{\rm n}$      & 1432.83$\pm$15.99 & 6562.77$\pm$0.01 & 0.44$\pm$0.01 \\
        6562 & H$\alpha_{\rm m}$      &   55.34$\pm$ 2.17 & 6562.66$\pm$0.02 & 1.29$\pm$0.02 \\
        6562 & H$\alpha_{\rm b}$      &   10.90$\pm$ 0.34 & 6564.23$\pm$0.23 & 8.87$\pm$0.34 \\
        6562 & H$\alpha^{\star}_{\rm n}$  &    7.79$\pm$ 0.52 & 6564.29$\pm$0.01 & 0.44$\pm$0.01 \\
        6583 & [NII]$_{\rm n}$        &   13.56$\pm$ 0.19 & 6583.41$\pm$0.01 & 0.33$\pm$0.01 \\
        6583 & [NII]$_{\rm m}$        &    1.66$\pm$ 0.13 & 6583.50$\pm$0.09 & 1.50$\pm$0.13 \\
        6678 & HeI$_{\rm n}$          &    9.84$\pm$ 1.65 & 6678.17$\pm$0.01 & 0.29$\pm$0.02 \\
        6678 & HeI$_{\rm m}$          &    5.42$\pm$ 1.63 & 6678.07$\pm$0.03 & 0.50$\pm$0.04 \\
        6716 & [SII]$_{\rm n}$        &   27.55$\pm$ 0.64 & 6716.43$\pm$0.01 & 0.33$\pm$0.01 \\
        6716 & [SII]$_{\rm m}$        &    3.92$\pm$ 0.48 & 6716.33$\pm$0.03 & 0.80$\pm$0.05 \\
        6730 & [SII]$_{\rm n}$        &   22.23$\pm$ 0.55 & 6730.81$\pm$0.01 & 0.33$\pm$0.01 \\
        6730 & [SII]$_{\rm m}$        &    3.27$\pm$ 0.40 & 6730.77$\pm$0.04 & 0.89$\pm$0.07 \\
        7002 & OI                     &    0.27$\pm$ 0.04 & 7002.20$\pm$0.04 & 0.28$\pm$0.04 \\
        7065 & HeI$_{\rm n}$          &   15.17$\pm$ 1.25 & 7065.20$\pm$0.01 & 0.35$\pm$0.01 \\
        7065 & HeI$_{\rm m}$          &    5.96$\pm$ 1.22 & 7065.28$\pm$0.02 & 0.64$\pm$0.04 \\
        7135 & [ArIII]$_{\rm n}$      &   23.40$\pm$ 1.07 & 7135.80$\pm$0.01 & 0.32$\pm$0.01 \\
        7135 & [ArIII]$_{\rm m}$      &    5.15$\pm$ 0.86 & 7135.69$\pm$0.04 & 0.73$\pm$0.06 \\
        7170 & [ArIV]                 &    0.36$\pm$ 0.07 & 7170.42$\pm$0.05 & 0.25$\pm$0.05 \\
        7254 & OI                     &    0.32$\pm$ 0.11 & 7254.09$\pm$0.09 & 0.24$\pm$0.09 \\
        7262 & [ArIV]                 &    0.34$\pm$ 0.12 & 7262.93$\pm$0.13 & 0.34$\pm$0.13 \\
        7281 & HeI                    &    3.09$\pm$ 0.09 & 7281.36$\pm$0.01 & 0.39$\pm$0.01 \\
        7319 & [OII]-a                &    5.32$\pm$ 0.27 & 7320.12$\pm$0.02 & 0.34$\pm$0.01 \\
        7319 & [OII]-b                &    2.58$\pm$ 0.25 & 7319.09$\pm$0.05 & 0.42$\pm$0.03 \\
        7330 & [OII]-a                &    2.84$\pm$ 0.20 & 7330.79$\pm$0.03 & 0.32$\pm$0.02 \\
        7330 & [OII]-b                &    3.56$\pm$ 0.21 & 7329.70$\pm$0.03 & 0.41$\pm$0.03 \\
        7377 & [NiII]                 &    0.31$\pm$ 0.03 & 7377.73$\pm$0.04 & 0.31$\pm$0.03 \\
        7411 & [NiII]                 &    0.09$\pm$ 0.03 & 7411.16$\pm$0.08 & 0.32$\pm$0.08 \\
        7468 & NI                     &    0.32$\pm$ 0.03 & 7468.31$\pm$0.04 & 0.35$\pm$0.04 \\
        7751 & [ArIII]                &    7.27$\pm$ 0.11 & 7751.09$\pm$0.01 & 0.38$\pm$0.01 \\
        7816 & HeI                    &    0.31$\pm$ 0.05 & 7816.13$\pm$0.06 & 0.36$\pm$0.06 \\
        8392 & Pa20                   &    1.51$\pm$ 0.08 & 8392.33$\pm$0.03 & 0.52$\pm$0.03 \\
        8413 & Pa19                   &    1.67$\pm$ 0.07 & 8413.27$\pm$0.02 & 0.51$\pm$0.02 \\
        8437 & Pa18                   &    1.84$\pm$ 0.07 & 8437.89$\pm$0.02 & 0.51$\pm$0.02 \\
        8446 & OI                     &    3.26$\pm$ 0.08 & 8446.39$\pm$0.01 & 0.47$\pm$0.01 \\
        8467 & Pa17                   &    2.34$\pm$ 0.05 & 8467.20$\pm$0.01 & 0.53$\pm$0.01 \\
        8502 & Pa16                   &    3.02$\pm$ 0.06 & 8502.45$\pm$0.01 & 0.52$\pm$0.01 \\
        8545 & Pa15                   &    3.38$\pm$ 0.16 & 8545.39$\pm$0.03 & 0.56$\pm$0.02 \\
        8598 & Pa14                   &    4.14$\pm$ 0.12 & 8598.38$\pm$0.02 & 0.54$\pm$0.02 \\
        8665 & Pa13                   &    4.92$\pm$ 0.10 & 8664.96$\pm$0.01 & 0.55$\pm$0.01 \\
        8680 & NI                     &    0.31$\pm$ 0.03 & 8680.18$\pm$0.05 & 0.46$\pm$0.05 \\
        8750 & Pa12                   &    5.84$\pm$ 0.13 & 8750.45$\pm$0.01 & 0.55$\pm$0.01\\
        8862 & Pa11                   &    7.93$\pm$ 0.14 & 8862.69$\pm$0.01 & 0.57$\pm$0.01 \\
        9014 & Pa10                   &   11.59$\pm$ 0.37 & 9014.83$\pm$0.02 & 0.54$\pm$0.01 \\
        9068 & [SIII]$_{\rm n}$       &   58.12$\pm$ 2.99 & 9068.89$\pm$0.02 & 0.44$\pm$0.02 \\
        9068 & [SIII]$_{\rm m}$       &    7.01$\pm$ 1.49 & 9068.79$\pm$0.07 & 1.12$\pm$0.12 \\

    \end{tabular}
    \caption{Best fit parameters of the model lines in the red arm.Column (1): theoric. Column (2): ion. Column (3): observed flux. Column (4): rest-frame wavelength [\AA]. Column (5): measured width [\AA]. The sub-indexes n, m and b correspond to the narrow, intermediate and broad components respectively. Lines marked with $\star$ denote the secondary red-shifted peak described in Section \ref{sec:multipleLines}. }
    \label{tab:lineFitsRED}
\end{table}

%%%%%%%%%%%%%%%%%%%%%%%%%%%%%%%%%%%%%%%%%%%%%%%%%%

% Don't change these lines
\bsp	% typesetting comment
\label{lastpage}
\end{document}